\documentclass[graybox, nosecnum]{svmult}

\UseRawInputEncoding
\usepackage{mathptmx}       % selects Times Roman as basic font
\usepackage{helvet}         % selects Helvetica as sans-serif font
\usepackage{courier}        % selects Courier as typewriter font
\usepackage{type1cm}        % activate if the above 3 fonts are
\usepackage{makeidx}         % allows index generation
\usepackage{graphicx}        % standard LaTeX graphics tool
\usepackage{multicol}        % used for the two-column index
\usepackage[bottom]{footmisc}% places footnotes at page bottom
\usepackage{hyperref}        %for hyperlinks
\usepackage{soul}            % for high-lighting of text
\hypersetup{colorlinks=true,urlcolor=blue}
\usepackage[square,numbers]{natbib}
\makeindex             % used for the subject index
\def\<<{{\ll}}
\def\>>{{\gg}}

\def\spose#1{\hbox to 0pt{#1\hss}}
\def\ltwig{\mathrel{\spose{\lower 3pt\hbox{$\mathchar"218$}}
     R_{\rm A}ise 2.0pt\hbox{$\mathchar"13C$}}}
\def\gtwig{\mathrel{\spose{\lower 3pt\hbox{$\mathchar"218$}}
     R_{\rm A}ise 2.0pt\hbox{$\mathchar"13E$}}}
\def\+/-{{\pm}}
\def\=={{\equiv}}

\def\ch{{\it Chandra}}
\def\xmm{{\it XMM-Newton}}

\newcommand{\beqa}{\begin{eqnarray}}
\newcommand{\eeqa}{\end{eqnarray}}

\newcommand{\Rkep}{\ensuremath{R_{\rm K}}}
\newcommand{\thooc}{$\theta^{1}$\,Ori\,C}
\newcommand{\sorie}{$\sigma$\,Ori\,E}

\newcommand{\tmdot}{\ensuremath{\tau_{\rm mass}}}
\newcommand{\Rstar}{\ensuremath{R_{\ast}}}
\newcommand{\Ralf}{\ensuremath{R_{\rm A}}}

\newcommand{\Mdot}{\ensuremath{\dot{M}}}
\newcommand{\estar}{\ensuremath{\eta_{\ast}}}

\newcommand{\etas}{\eta_\ast}
\newcommand{\ra}{R_\mathrm{A}}
\newcommand{\rk}{R_\mathrm{K}}
\newcommand{\rs}{R_\ast}
\newcommand{\teff}{T_\mathrm{eff}}

\newcommand{\tauj}{\tau_\mathrm{J}}
\newcommand{\tsmax}{t_{\mathrm{s,max}}}
\begin{document}
%\tableofcontents{}
\title*{Magnetically confined wind shock }
% Use \titlerunning{Short Title} for an abbreviated version of
% your contribution title if the original one is too long
\author{Asif ud-Doula \thanks{corresponding author} and Stan Owocki}
% Use \authorrunning{Short Title} for an abbreviated version of
% your contribution title if the original one is too long
\institute{Asif ud-Doula \at Penn State Scranton, 120 Ridge View Drive, Dunmore, PA 18512, USA, \email{asif@psu.edu}
\and Stan Owocki \at University of Delaware, Newark, DE 19711, USA \email{owocki@bartol.udel.edu}}
%
% Use the package "url.sty" to avoid
% problems with special characters
% used in your e-mail or web address
%
\maketitle
\abstract{Many  stars across all classes possess strong enough magnetic fields to influence dynamical flow of material off the stellar surface. 
For the case of massive stars (O and B types),  about 10\% of them harbour strong, globally ordered (mostly dipolar) magnetic fields. The trapping and channeling of their stellar winds in closed magnetic loops leads to {\it magnetically confined wind shocks} (MCWS), with pre-shock flow speeds that are some fraction of the wind terminal speed that can be a few thousand km s$^{-1}$.  These shocks generate hot plasma, a source of X-rays. In the last decade, several developments took place, notably the determination of the hot plasma properties for a large sample of objects using \xmm\ and \ch, as well as fully self-consistent MHD modelling and the identification of shock retreat effects in weak winds. In addition, these objects are often sources of H$\alpha$ emission which is controlled by either sufficiently high  mass loss rate or  centrifugal breakout. Here we review the theoretical  aspects of such magnetic massive star wind dynamics. 
}
\section{Keywords} 
Massive stars; magnetic field; rotation; x-rays; stellar winds, radiative cooling.
\section{Introduction}
Hot luminous, massive stars of spectral type O and B are prominent sources of X-rays which can  originate from three distinct sources: shocks in their high-speed radiatively driven stellar winds, wind-wind collisions in binary systems and magnetically confined wind shocks. 

In  single, non-magnetic O stars, the intrinsic instability of wind driving by line-scattering leads to embedded wind shocks that are thought to be the source of  their relatively soft ($\sim$0.5\,keV) X-ray spectrum, with a total X-ray luminosity that scales with stellar bolometric luminosity, $L_{\rm x} \sim 10^{-7} \times L_{\rm bol}$  \cite{Chlebowski89, Berghoefer97, Naze11}. In  massive binary systems the collision of the two stellar winds at up to the wind terminal speeds can lead to even higher $L_{\rm x}$, generally with a significantly harder (up to 10\,keV)  spectrum.

Here we discuss a third source of X-rays from OB winds, namely those observed from the subset ($\sim$10\%) of massive stars with strong, globally ordered (often significantly dipolar) magnetic fields \cite{Petit13};  in this case, the trapping and channeling of the stellar wind in closed magnetic loops leads to {\em magnetically confined wind shocks}  (MCWS) \cite{BabMon1997a,BabMon1997b}, with pre-shock flow speeds that are some fraction of the wind  terminal speed, resulting in intermediate energies for the shocks and associated X-rays  ($\sim$2\,keV). A prototypical example is provided by the magnetic O-type  star $\theta^1$~Ori~C, which shows moderately hard X-ray emission with a rotational phase variation that matches well the expectations of the MCWS paradigm  \cite{Gag2005}.

Here, we focus on theoretical aspects of magnetic confinement that determine the extent of the influence of the field over the wind while the observational aspects are addressed in another chapter titled `X-ray emission of massive stars and their winds'. We, then, describe an effect called `shock retreat', which can moderate the strength of the X-rays, or even quench it altogether in extremely low mass loss rate stars. 
We also review the properties and observational signatures (e.g.  Balmer line emission) of the resulting  magnetospheres in closed loop regions, and on the stellar rotation spindown that results from the angular momentum loss associated with magnetically torqued wind outflow from open field regions. In this way magnetic fields can have a profound effect on the starÕs rotational evolution, giving rotation periods ranging from weeks to even decades, in strong contrast to the day-timescale periods of non-magnetic massive stars.

\section{\textit{Historical Perspective}}
To explain X-ray emission from the  Ap/Bp star IQ Aur  \cite{BabMon1997a} introduced  the MCWS model. In their approach, they effectively prescribed a fixed magnetic field geometry to channel the wind outflow (see also \cite{ShoBro1990}). For large magnetic loops, wind material from opposite footpoints is accelerated to a substantial fraction of the wind terminal speed (i.e., $\ge$1000 km s$^{-1}$) before the channeling toward the magnetic field loop tops forces a collision with very strong shocks, thereby heating the gas to temperatures (10$^7$ - 10$^8$ K) that are high enough to emit hard (few keV) X-rays. This star has a quite strong field ($\sim$4 kG) and a rather weak wind, with an estimated mass loss rate of about $\sim 10^{-10} M_\odot$ yr$^{-1}$, and thus  indeed could be reasonably modeled within the framework of prescribed magnetic field geometry. However, the actual X-ray emission from the star, which is predominantly soft, is further influenced by an effect called `shock retreat' that we describe below.  Later,   \cite{BabMon1997b} applied this model to explain the periodic variation of X-ray emission of the O7 star $\theta^1$~Ori~C, which has a lower magnetic field ($\sim$1100 G) and significantly stronger wind (mass-loss rate $\sim 10^{-7} M_\odot$ yr$^{-1}$), raising now the possibility that the wind itself could influence the field geometry in a way that is not considered in the simple fixed-field approach.

\subsection{Magnetic Confinement}
In an interplay between magnetic field and stellar wind,
the dominance of the field is determined  by how strong it
is relative to the wind. To understand the
competition between these two,  \cite{udDOwo2002} defined a characteristic
parameter for the relative effectiveness of the magnetic fields in
confining and/or channeling the wind outflow. Specifically, consider
the ratio between the energy densities of field vs. flow,
\begin{eqnarray}
\eta (r, \theta) &\equiv& \frac{B^2/8\pi}{\rho v^2/2} 
\approx \frac{B^2 r^2}{\dot{M}v(r)} \label{etadef}
\\
&=& \left [\frac{B_{\ast}^2 (\theta )
{R_{\ast}}^2}{\dot{M}v_{\infty}}\right ] \left
[\frac{(r/R_{\ast})^{-2n}}{1-R_{\ast}/r} \right ] \, , \nonumber
\end{eqnarray}
where the latitudinal variation of the surface field has the dipole
form given by $B_\ast^2 (\theta) = B_o^2 ( \cos^2 \theta + \sin^2
\theta/4 )$.
In general, a magnetically channeled outflow will have a complex
flow geometry, but for convenience, the second equality in eqn.~(\ref{etadef}) simply characterizes the wind  strength in terms of a
spherically symmetric mass loss rate $\dot{M}=4\pi r^2 \rho v$. The
third equality likewise characterizes the radial variation of
outflow velocity in terms of the phenomenological velocity law $v(r)
=v_{\infty}(1-R_{\ast}/r)^\beta$, with $v_{\infty}$ the wind terminal
speed and assumed $\beta=1$ which seems to well approximate the numerical solution of massive star winds. This equation furthermore models the magnetic field strength
decline as a power-law in radius, $B(r)
=B_{\ast}(R_{\ast}/r)^{(n+1)}$, where, e.g., for a  dipole
$n=2$.

With the spatial variations of this energy ratio thus isolated
within the right square bracket, we see that the left square bracket
represents a dimensionless constant that characterizes the overall
relative strength of field vs. wind. Evaluating this in the region
of the magnetic equator ($\theta=90^o$), where the tendency toward a
radial wind outflow is in  most direct competition with the tendency
for a horizontal orientation of the field, one can thus define an
equatorial `wind magnetic confinement parameter',
\begin{eqnarray}
\eta_{\ast} &\equiv&
\frac{B_\ast^2 (90^\circ) {R_{\ast}}^2} {\dot{M}v_{\infty}}
= 0.4  \,  \frac{B_{100}^2 \, R_{12}^2}{\dot{M}_{-6} \, v_8}.
\label{wmcpdef}
\end{eqnarray}
where $\dot{M}_{-6} \equiv \dot{M}/(10^{-6}\, M_{\odot}$/yr),
$B_{100} \equiv B_o/(100$~G), $R_{12} \equiv R_{\ast}/(10^{12}$~cm),
and $v_{8} \equiv v_{\infty}/(10^8$~cm/s). 
 In order to have any confinement, $\eta_\ast \ge 1$.
As these stellar and wind
parameters are scaled to typical values for an OB supergiant, e.g.
$\zeta$ Pup, the last equality in eqn. (\ref{wmcpdef}) immediately
suggests that for such winds, significant magnetic confinement or
channeling should require fields of order  few hundred G. By
contrast, in the case of the Sun, the much weaker mass loss (${\dot
M}_\odot \sim 10^{-14}~M_{\odot}$/yr) means that even a much weaker
global field ($B_{o} \sim 1$~G) is sufficient to yield $\eta_{\ast}
\simeq 40$, implying a substantial magnetic confinement of the solar
coronal expansion. But in Bp stars the magnetic field strength can be of order kG with 
${\dot M}_\odot \sim 10^{-10}~M_{\odot}$/yr leading $\eta_{\ast} \le 10^6$. 
Thus, the confinement in Bp stars is very extreme.

We emphasize that  $\dot{M}$ used in the above 
formalism is a value obtained for a spherically symmetric non-magnetic wind  as the magnetic field may significantly influence the predicted circumstellar density and velocity structure. 

\subsubsection{Alf\'ven Radius}
The extent of the effectiveness of magnetic  confinement
is set by the Alfv\'en radius, $R_A$, where flow and Alfv\'en velocities are equal. This will also determine
the extent of the largest loops and thus the highest shock velocities affecting the hardness of X-ray emission.
This radius can be derived from eqn.~(\ref{etadef}) where the second square bracket factor
shows the overall radial variation;
$n$ is the power-law exponent for radial decline of
the assumed stellar field, e.g. $n=2$ for a pure dipole.
For a star with a non-zero field, we have $\eta_{\ast} > 0$, and so given
the vanishing of the flow speed at the atmospheric wind base, this
energy ratio always starts as a large number near the stellar surface,
$\eta(r \rightarrow R_{\ast}) \rightarrow \infty$.
But from there outward it declines quite steeply, asymptotically as
$r^{-4}$ for a dipole, crossing unity at the
Alfv\'{e}n radius defined implicitly by  $\eta(R_A) \equiv 1$ leading to :
\begin{equation}
1 = \eta_\ast
\left [ \frac{(R_A/R_{\ast})^{-2n}}{1-R_{\ast}/R_A} \right ] 
\end{equation}

Thus, for a canonical $\beta=1$ wind velocity law,
explicit  solution for $R_{\rm{A}}$ along the magnetic equator
requires finding the appropriate root of
\begin{equation}
    \left ( {R_A \over R_\ast } \right )^{2n} -
    \left ( {R_A \over R_\ast } \right )^{2n-1} = \eta_{\ast}
\, ,
\label{radef}
\end{equation}
which for integer $2n$ is just a simple polynomial, specifically
a quadratic, cubic, or quartic for $n =$~1, 1.5, or 2.
Even for non-integer values of $2n$, the relevant solutions
can be approximated (via numerical fitting)
to within a few percent by the simple general expression,
\begin{equation}
\frac{R_{\rm{A}}}{R_{\ast}}
\approx 1  + (\eta_{\ast} + 1/4)^{1/(2n)} -  (1/4)^{1/(2n)}
\, .
\label{raapp}
\end{equation}
For weak confinement, $\eta_{\ast} \ll 1$, we find
$R_{\rm{A}} \rightarrow R_{\ast}$,
while for strong confinement, $\eta_{\ast} \gg 1$, we obtain
$R_{\rm{A}} \rightarrow \eta_{\ast}^{1/(2n)} R_{\ast}$.
In particular, for the standard dipole case with $n=2$,
we expect the strong-confinement scaling
$R_{\rm{A}}/R_{\ast} \approx \eta_{\ast}^{1/4}$.

Clearly $R_{\rm{A}}$ represents the radius at which the wind speed $v$
exceeds the local Alfv\'{e}n speed $V_{A}$.
It also characterizes the maximum radius where the
magnetic field still dominates over the wind. For  Ap/Bp stars where stellar fields are of order kG,
$\eta_\ast \gg 1$, e.g. for $\sigma$~Ori~E it is about $10^7$,
implying an Alfv\`{e}n radius $\sim 60 R_\ast$.  Thus, in Bp (and Ap) stars wind is trapped
to large radii creating extensive magnetospheres. This also implies that X-rays from Bp stars should be intrinsically hard.  But as
we show below, shock retreat effects may soften it significantly.

\subsection {Rotation and Kepler Radius}

\begin{figure*}
\begin{center}
	\includegraphics[width=80mm]{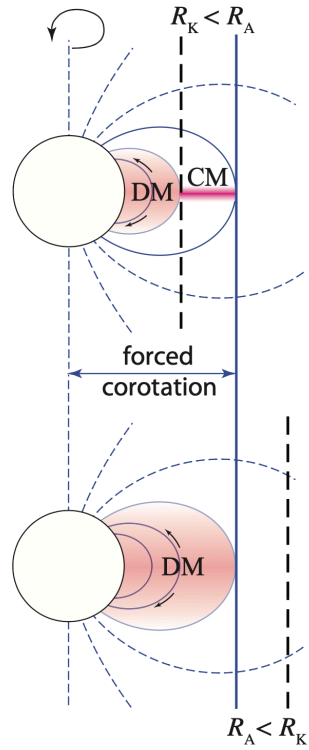}
%\vspace{-0.7in}
\caption{\small{
Sketch of the regimes for a dynamical vs.\ centrifugal magnetosphere (DM vs. CM).
}
\label{fig:dmcm} 
}
\end{center}
\end{figure*}

Another important parameter, rotation, can have dynamical effects and can be  parameterized (see \cite{Uddoula08}) in terms of the orbital rotation fraction,

 \begin{equation}
 \label {eq-W}
  W \equiv \frac{V_{rot}}{V_{orb}} = \frac{V_{rot}}{\sqrt{GM_\ast/R_\ast}} \, 
    \end{equation}
  where $V_{rot}$ is the star`s equatorial rotation speed and $V_{orb}$ is the Keplerian orbital speed near the equatorial surface radius $R_\ast$. 
 Insofar as the field within the Alfv\`en radius is strong enough to maintain rigid-body rotation, the Kepler corotation radius $R_K$  identifies where the centrifugal force for rigid-body rotation exactly balances the gravity in the equatorial plane. If $R_A < R_K$, then material trapped in closed loops will again eventually fall back to the surface, forming a \emph {dynamical magnetosphere} (DM). But if $R_A > R_K$, then wind material located between $R_K$ and $R_A$ can remain in static equilibrium, forming a \emph{centrifugal magnetosphere }(CM) that is supported against gravity by the magnetically enforced co-rotation. This  is illustrated by  figure \ref{fig:dmcm}.  We can then compute $R_K$ readily from:
  \begin{equation}
 \label {eq-Rk}
  R_K=W^{-2/3} R_\ast .
  \end{equation}

\begin{figure*}
\begin{center}
\vfill
\includegraphics[width=38mm]{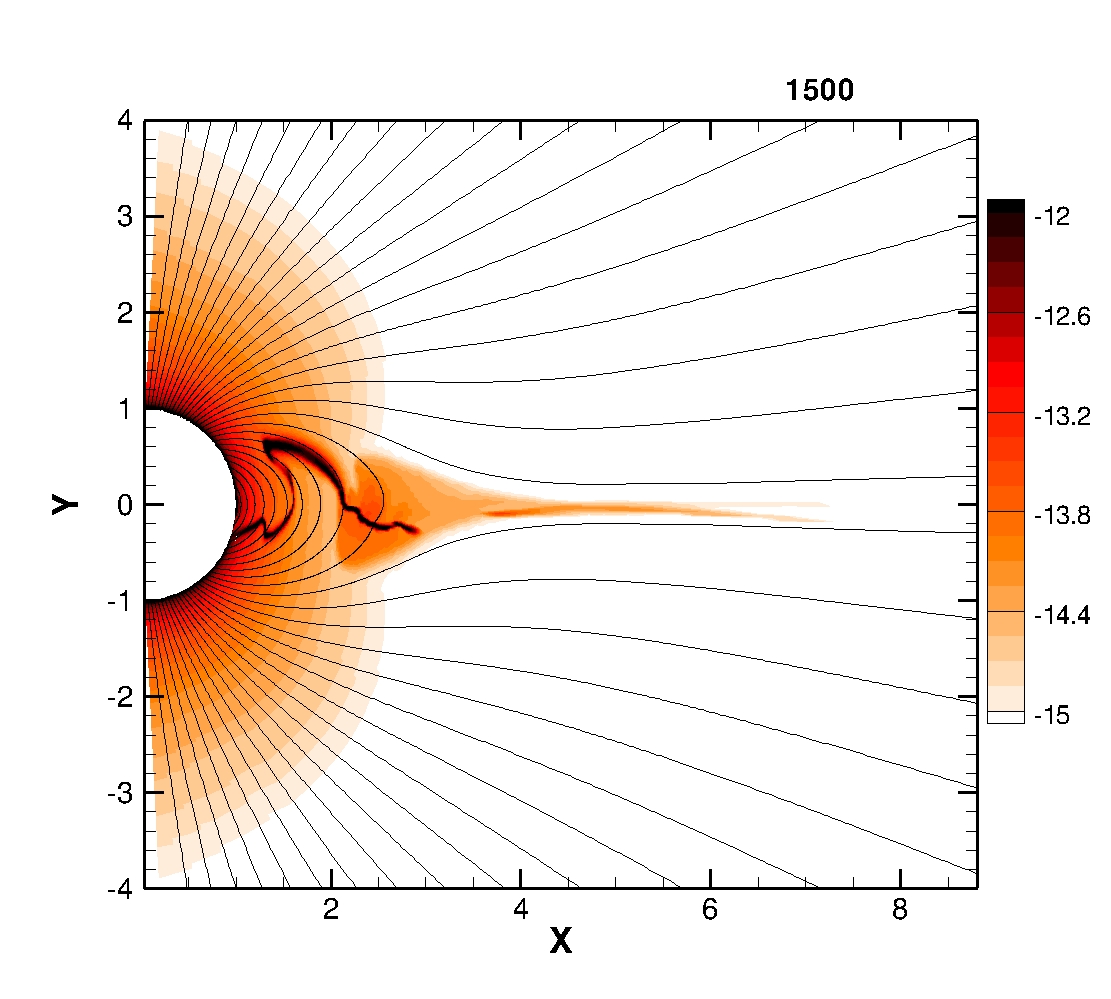}
\includegraphics[width=38mm]{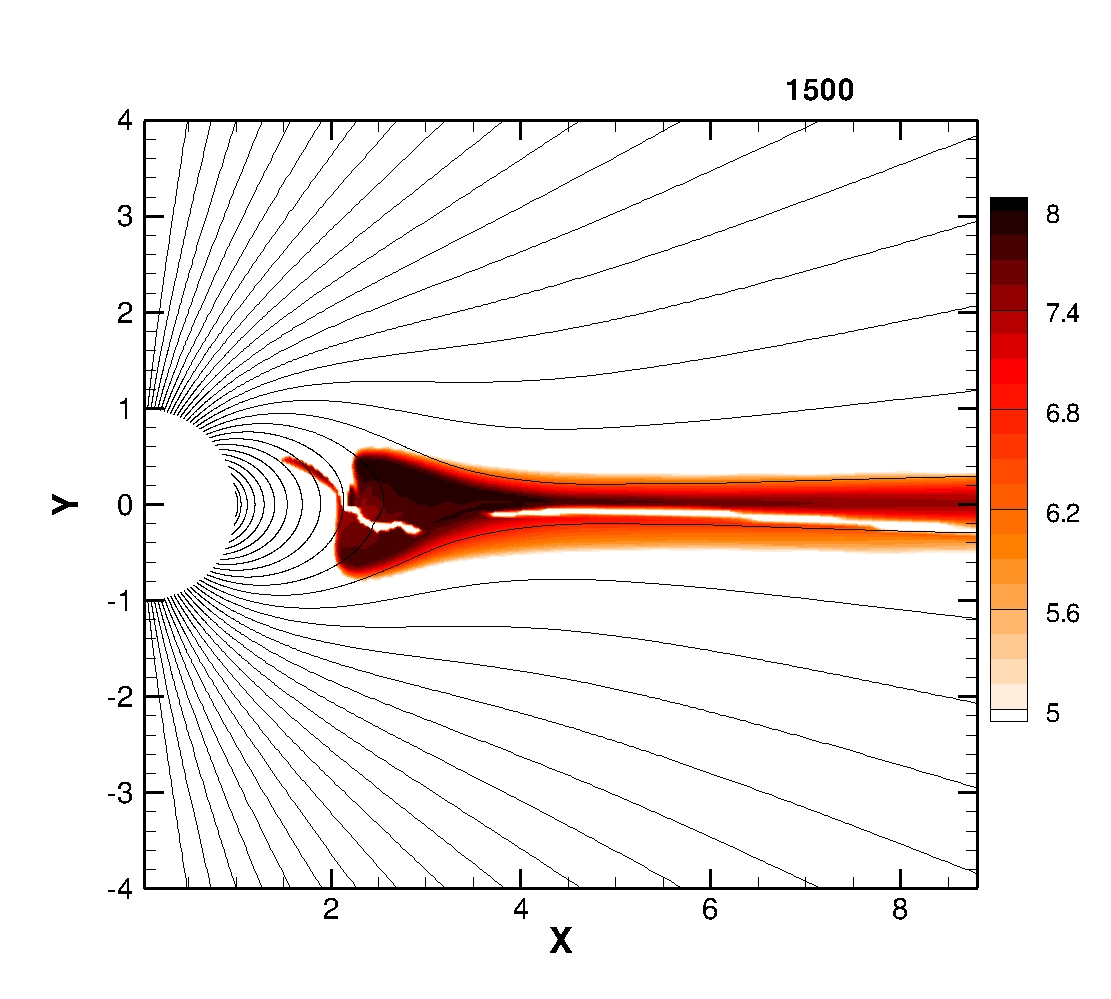}
\includegraphics[width=38mm]{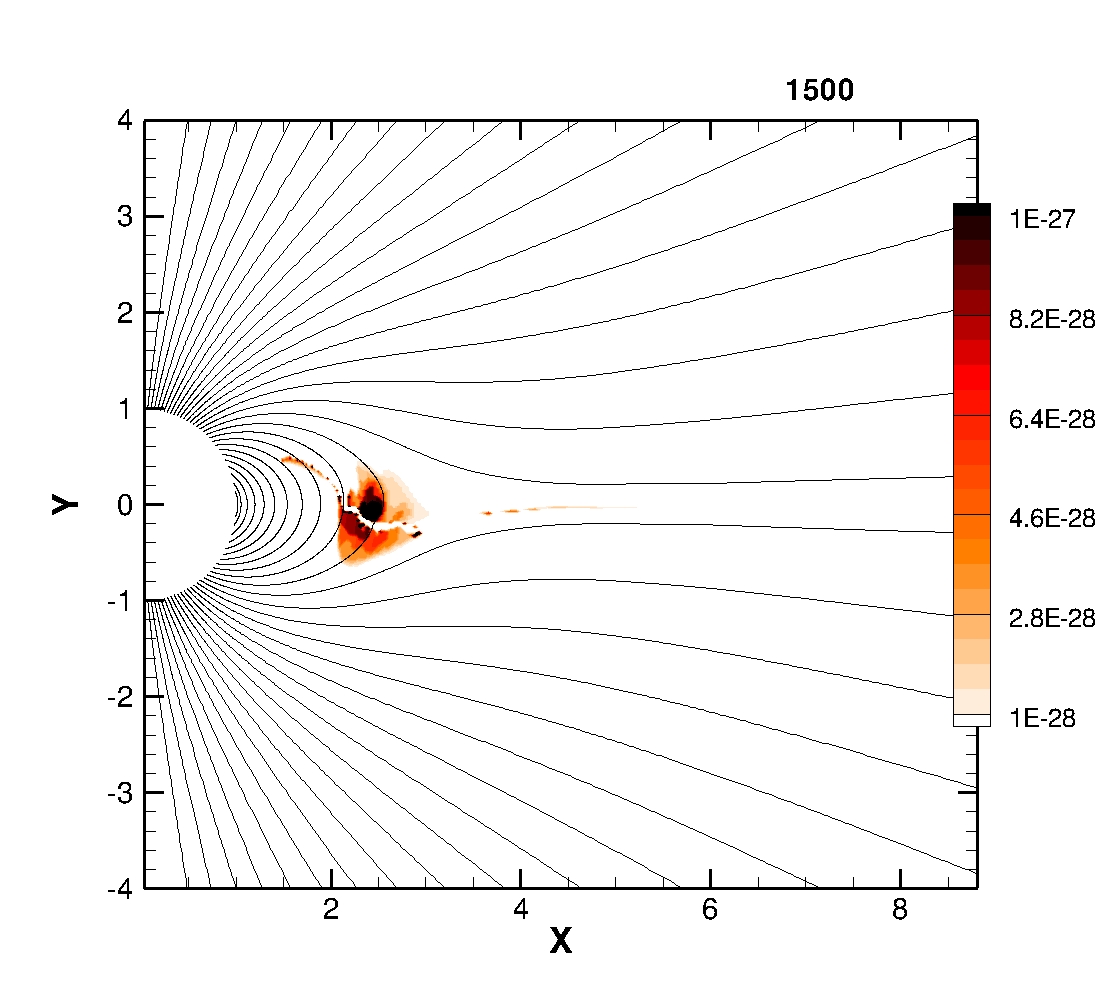}
\caption{
For a 2D MHD simulation of a magnetized wind with confinement parameter $\eta_\ast = 100$,
(\cite{Uddoula14})
 color plots of log density (left) and log temperature (middle) for arbitrary snapshot many dynamical times after initialization.
 Note that magnetic loops for  initially a dipole field extending above $\ra/\rs \approx 100^{1/4} \approx 3.2$ are drawn open by the wind, while those with an apex below $\ra$ remain closed.
The right panel plots associated X-ray emission from the magnetically confined wind shock (MCWS) near the apex of closed loops. 
}
\label{fig:rhotxem}
\end{center}
\end{figure*}

\begin{figure}
\begin{center}
	\includegraphics[width=115mm]{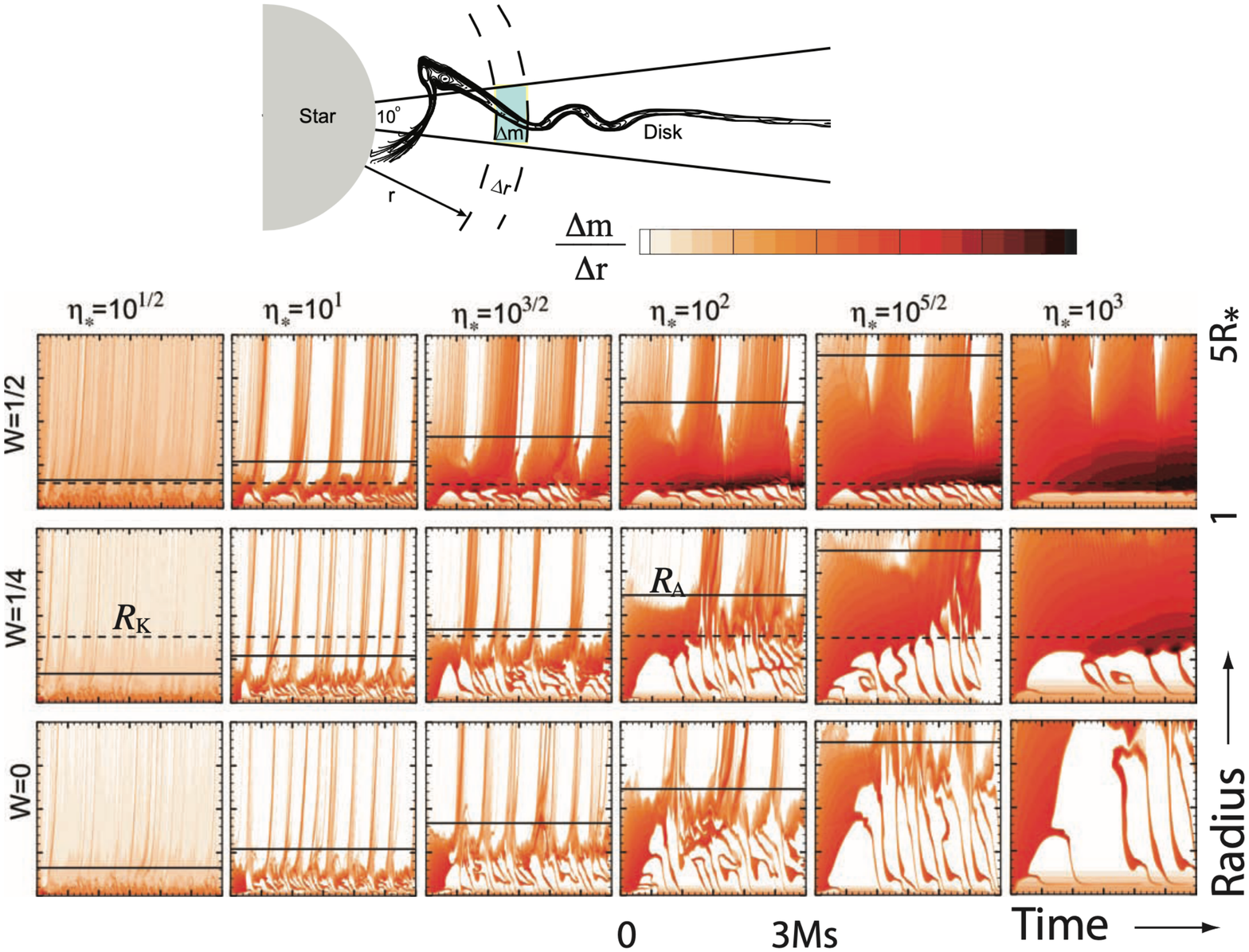}
%\vspace{-0.7in}
\caption{\small{
{\em Top:}
 Contour plot for density at an arbitrary snapshot of an isothermal 2D MHD simulation, overlayed with illustration to define the radial mass distribution, $\Delta m/\Delta r$ near the equator.  
{\em Bottom:}
Color plots for  log of $\Delta m/\Delta r$, plotted versus radius (1-5 \Rstar) and time (0-3~Msec), for a mosaic of 2-D MHD models with a wide range of magnetic confinement parameters \estar, and 3 orbital rotation fractions $W$. 
The horizontal solid lines indicate the Alfv\'en radius \Ralf\ (solid) and Kepler radius \Rkep\ (dashed).
}
\label{fig:dmdr} 
}
\end{center}
\end{figure}

  \subsubsection{MHD Simulations}

The initial magnetohydrodynamic (MHD) simulations by  \cite*{udDOwo2002} assumed, for
simplicity, that radiative heating and cooling would keep the wind
outflow nearly isothermal at roughly the stellar effective temperature. 
The simulations studied the dynamical competition
between field and wind by evolving  MHD simulations
from an initial condition when a dipole magnetic field is suddenly introduced into a previously relaxed,
one-dimensional spherically symmetric wind.

Immediately after the introduction of the field, the dynamic interplay between the wind and the field leads to 
two distinct regions. Along the polar region, the wind freely streams radially outward, stretching the field lines into a radial configuration, as can 
be inferred from the left panel of illustrative Fig. \ref{fig:rhotxem}. If the field is strong enough, around the magnetic equator a region of closed magnetic
loops is formed wherein the flow from opposite hemispheres collides to make strong shocks, quite similar to what was predicted in the semi-analytic, fixed-field models of 
\cite{BabMon1997a}.  The shocked material forms a dense disk-like structure which is opaque to line-driving. 
But its support against gravity by the magnetic tension along the convex field lines is inherently unstable, leading to a complex pattern of fall back along the loop lines down to the star,  again as suggested by the left panel of  Fig.  \ref{fig:rhotxem} .

Note that even for weak field models with moderately small confinement, $\eta_{\ast} \le 1/10$,  the field still has a noticeable global influence
on the wind, enhancing the density and decreasing the flow speed near
the magnetic equator. However, shock speeds are probably not  large enough to produce any X-rays in that case.

The mosaic of color plots in figure \ref{fig:dmdr} shows the time vs. height variation of the equatorial mass distribution $\Delta m/ \Delta r$ for various combinations of rotation fraction $W$ and wind confinement $\estar$.
Note the DM infall for material trapped below $\Rkep$ and $\Ralf$, vs.\ the dense accumulation of a CM from confined material near and above $\Rkep$, but below $\Ralf$.

This diagram thus provides a vivid way to characterize how the global geometry and time evolution of magnetospheres of massive stars (MSMS) depends on rotation and magnetic confinement.
However, it is based entirely on the special case of field-aligned rotation in 2D.

To model the actual X-ray emission from shocks that form from  the magnetic
channeling  and confinement, subsequent efforts \cite[][]{udD2003, Gag2005}
have relaxed  the assumption of isothermal equation of state in earlier studies to include a detailed energy equation
that follows the radiative cooling of shock-heated material.
The MCWS model provides excellent agreement with the diagnostics from the phase-resolved {\it Chandra} spectroscopy of $\theta^1$~Ori~C \citep{Gag2005}.

\subsubsection{Rotation-confinement diagram and stellar spindown}

\begin{figure}[t!]
%\begin{figure}
{
	\includegraphics[width=115mm]{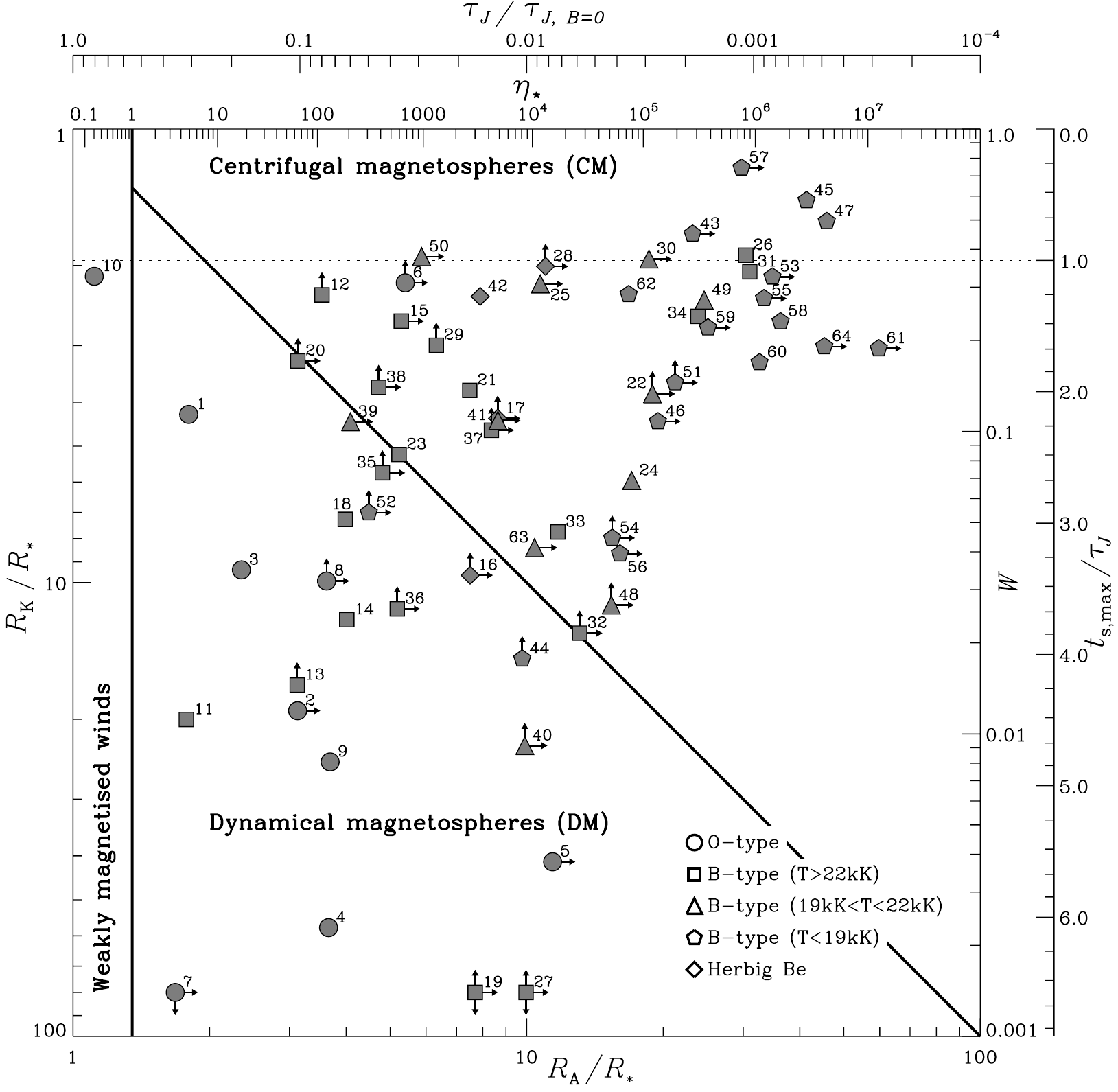}	
}
\caption{\small{
Classification of observationally confirmed magnetic massive stars in terms of  magnetic confinement vs.\ rotation fraction, characterized here by a log-log plot of Kepler radius $\Rkep$ increasing downward vs.\ Alfv\'{e}n radius $\Ralf$  increasing to the right
\citep{Petit13}. 
The diagonal line  separates the domains of dynamical magnetospheres (DM) with $\Ralf<\Rkep$ vs.\ centrifugal magnetospheres (CM) with $\Ralf>\Rkep$.
The additional upper and right axes give respectively the corresponding spindown timescale $\tauj$ 
and maximum spindown age $\tsmax$ 
(see equation \ref{eq:tbrake} and following text).
Rapidly rotating stars above the horizontal dotted line have $\tsmax < \tauj$.
}
 }
\label{fig:ipod}
\end{figure}

This theoretical characterization of the properties of MSMS provides an 
instructive way to classify their observational characteristics within a {\em rotation vs. magnetic confinement diagram}.
For the still-growing list of observationally confirmed magnetic hot-stars (with $\teff \ge 16$\,kK) compiled by \citep{Petit13}, the left panel of figure \ref{fig:ipod} plots positions in a log-log plane of $\Rkep$ vs. $\Ralf$.
The diagonal line representing $\Rkep = \Ralf$ divides the domain of 
CMs  to the upper right from that for 
DMs to the lower left. Note that almost all {\em all} O-stars 
are  located among the slow rotators in the lower left, a direct consequence of their rapid spindown from the loss of angular momentum in their strong, magnetized winds.

In DMs,  all the  material trapped in closed magnetic loops eventually falls back to the star, and so balances the input from the upflowing stellar wind.  
This can greatly reduce the overall net mass loss, implying a higher final remnant mass in the stellar evolution that could help explain the large masses inferred 
from LIGO gravitational wave detections of black hole mergers \citep{Petit17}.
Within such closed loops, the overall structure of shocked-heated gas, followed by cooling and infall, is well characterized by an {\em analytic dynamical magnetosphere} \citep[ADM;][]{Owocki16} model. 
As discussed in this chapter, this generalized ADM model has proven quite useful for interpreting general trends in observational diagnostics like X-ray emission \citep{Uddoula14},
%(ud-Doula et al.\ 2014), 
UV wind line variations \Citep{Erba19}, and H-alpha emission \citep{Sundqvist12}. 

2D MHD aligned-dipole  simulations \citep{Uddoula09}  have indicated that the angular momentum carried out by a magnetically torqued stellar wind follows the same simple, split-monopole scaling law derived for the Sun by \citep{Weber67}, $\dot J =\frac{2}{3} {\dot M} \, \Omega \, \Ralf^{2}$ -- with, however, the Alfv\'en radius \Ralf\ now given by the {\em dipole} scaling $\Ralf\sim \estar^{1/4}$ (see equation \ref{raapp} for $\eta_\ast \gg 1$ and $n=2$), instead the oft-quoted, stronger scaling ($\Ralf\sim \estar^{1/2}$) for a split monopole. This leads to an associated general  formula for the rotational spindown timescale,
\begin{equation} \label{eq:tbrake}
     \tau_{\rm J} \equiv \frac{I \Omega}{{\dot J}}
     = \frac{3}{2} f \tmdot \, \left ( \frac{\Rstar}{\Ralf}\right )^{2}
     \approx 0.15 \frac{\tmdot}{\sqrt{\estar}} 
     \, ,
\end{equation}
where $\tmdot \equiv M/{\dot M}$ is the stellar mass loss timescale, and $f $ is a dimensionless measure of the star's moment of inertia 
$I \equiv f M\Rstar^2$. For the case of a rigid sphere $f \equiv \frac{2}{5}$, but  massive stars with convective cores and radiative envelopes rarely behave like a rigid sphere, and  assumed value $f \approx 0.1$  in the above equation is a good approximation.

This can be used to define a star's {\em maximum spindown age} $\tauj$ (i.e., the number of spindown e-folds from an assumed critical initial rotation)  in terms of its inferred present-day critical rotation fraction $W$ relative to an assumed initial critical rotation $W_o =1$,
$\tsmax = {\tauj} (\ln{W_o/W} )$.
In  figure \ref{fig:ipod} the upper axis gives the spindown timescale $\tauj$ (normalized by the value in a non-magnetized wind), 
while the right axis gives the maximum spindown age $\tsmax$ 
(normalized by the spindown time).
Stars above the horizontal dotted line have a maximum spindown age that is {\em less} than a single spindown time.
All the most rapidly rotating stars are cooler B-type with weak winds, and thus weak braking, despite their strong field.
This spindown scaling thus forms the basis for modeling the rotational evolution of magnetic massive stars \citep{Keszthelyi20}. It explains why nearly all the magnetic O-stars are slow rotators.
Moreover, application to the prototypical magnetic B-star  \sorie\  shows good agreement with the spindown  directly inferred from extended monitoring of the timing of magnetospheric clouds transiting in front of the star \citep{Tow2010}.

Despite this general success, a key open question regards how this angular momentum loss scaling, which is based on 2D simulations  of field-aligned dipole cases, might be altered by 3D effects for tilted dipoles or higher-order multipoles. Current preliminary results from 3D MHD simulations indicate that the effects are  of order unity.

\begin{figure}[t!]
\vspace{-0.0in}
\begin{center}
\includegraphics[scale=0.275]{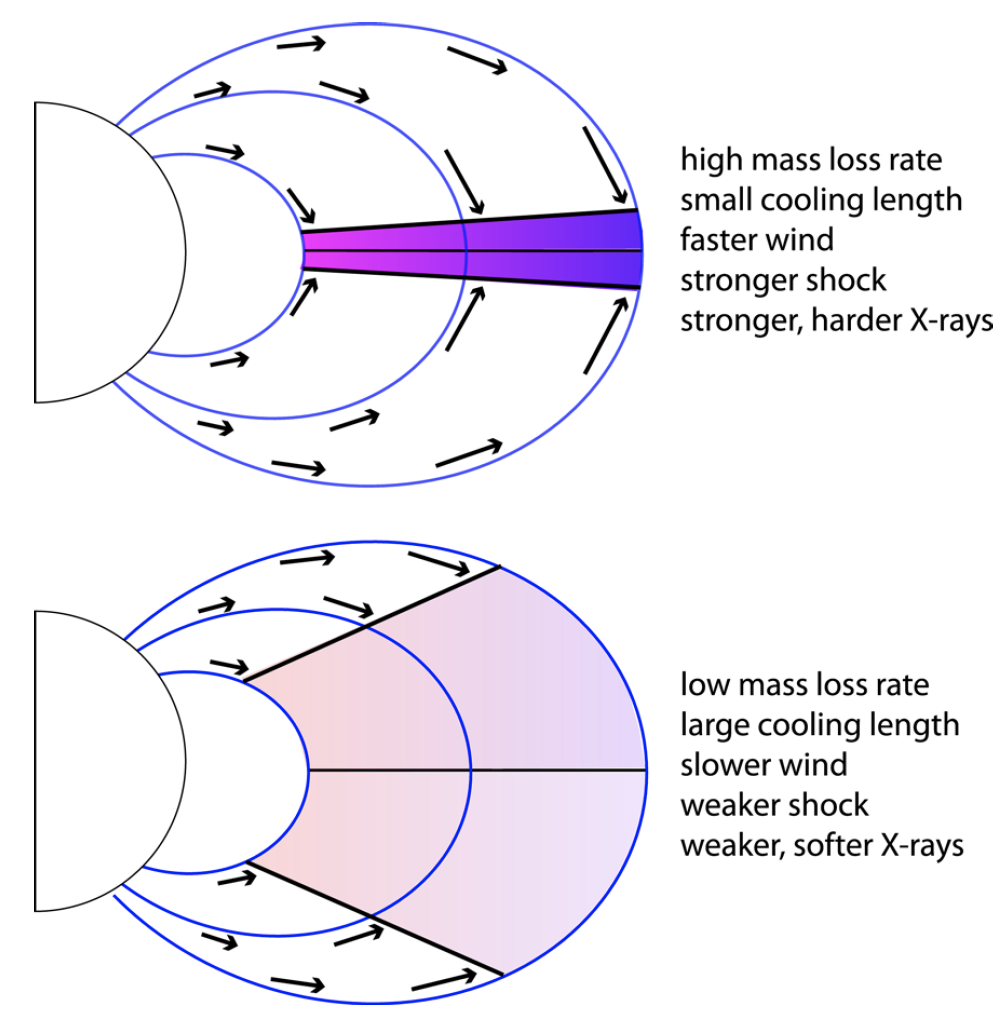}
\includegraphics[scale=0.245]{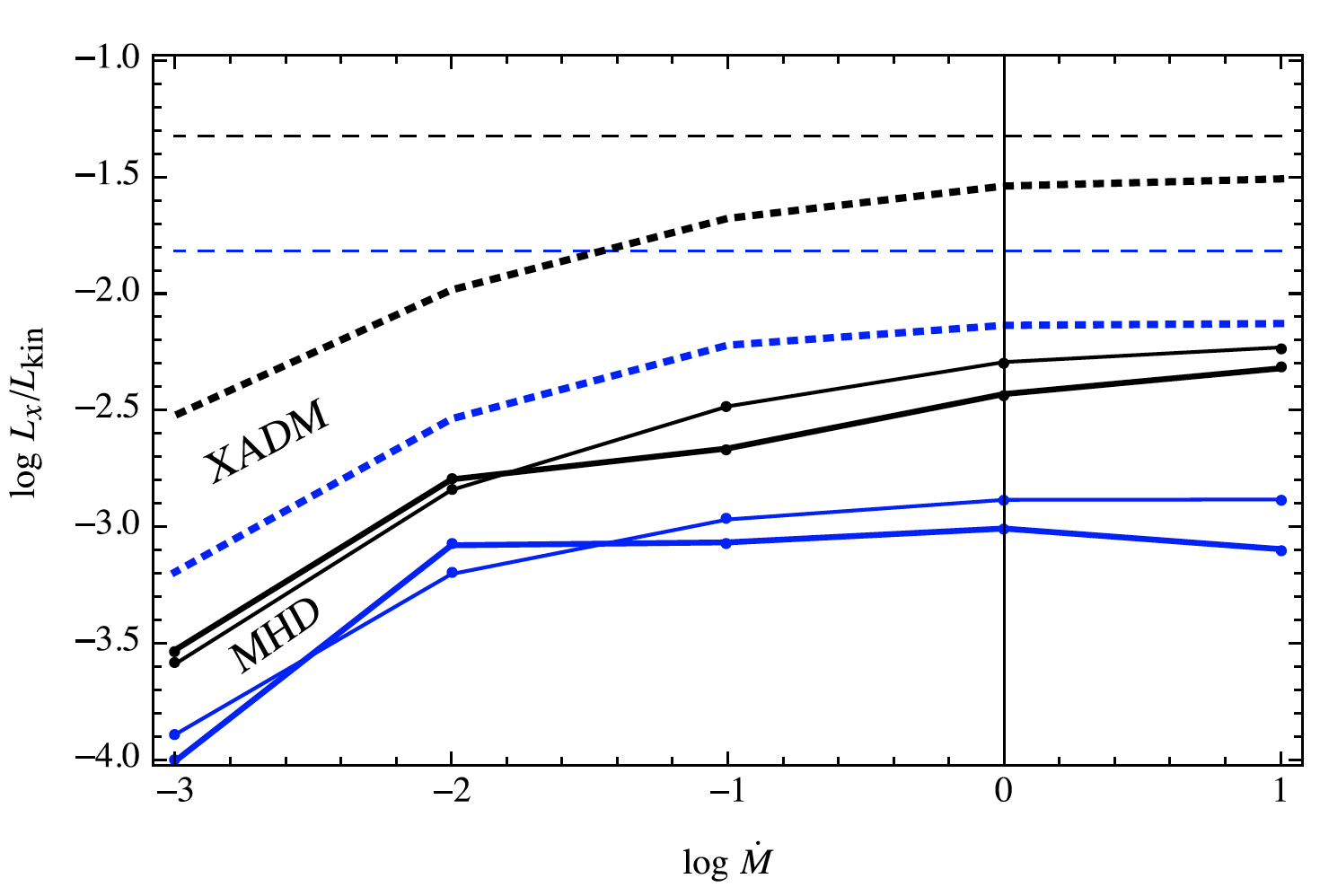}
\includegraphics[scale=0.31]{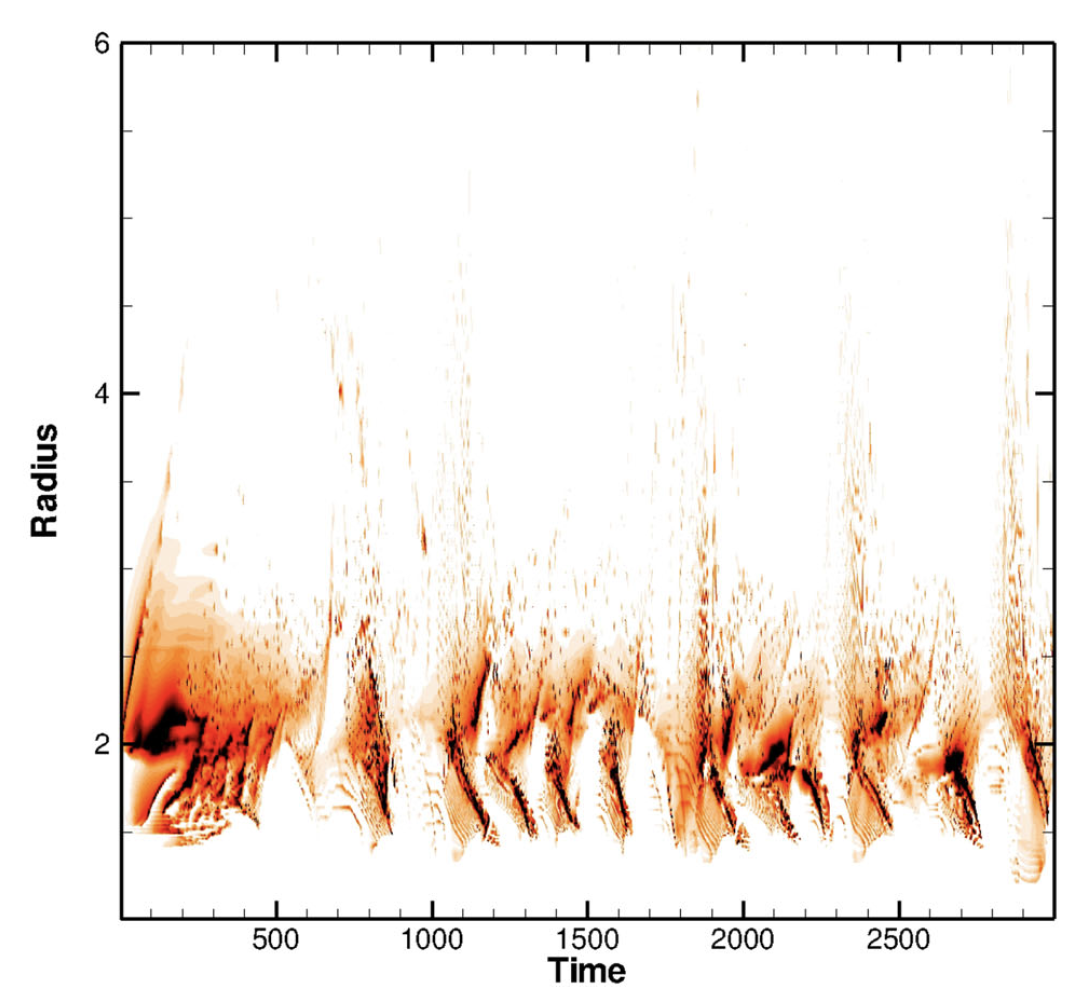}
\includegraphics[scale=0.245]{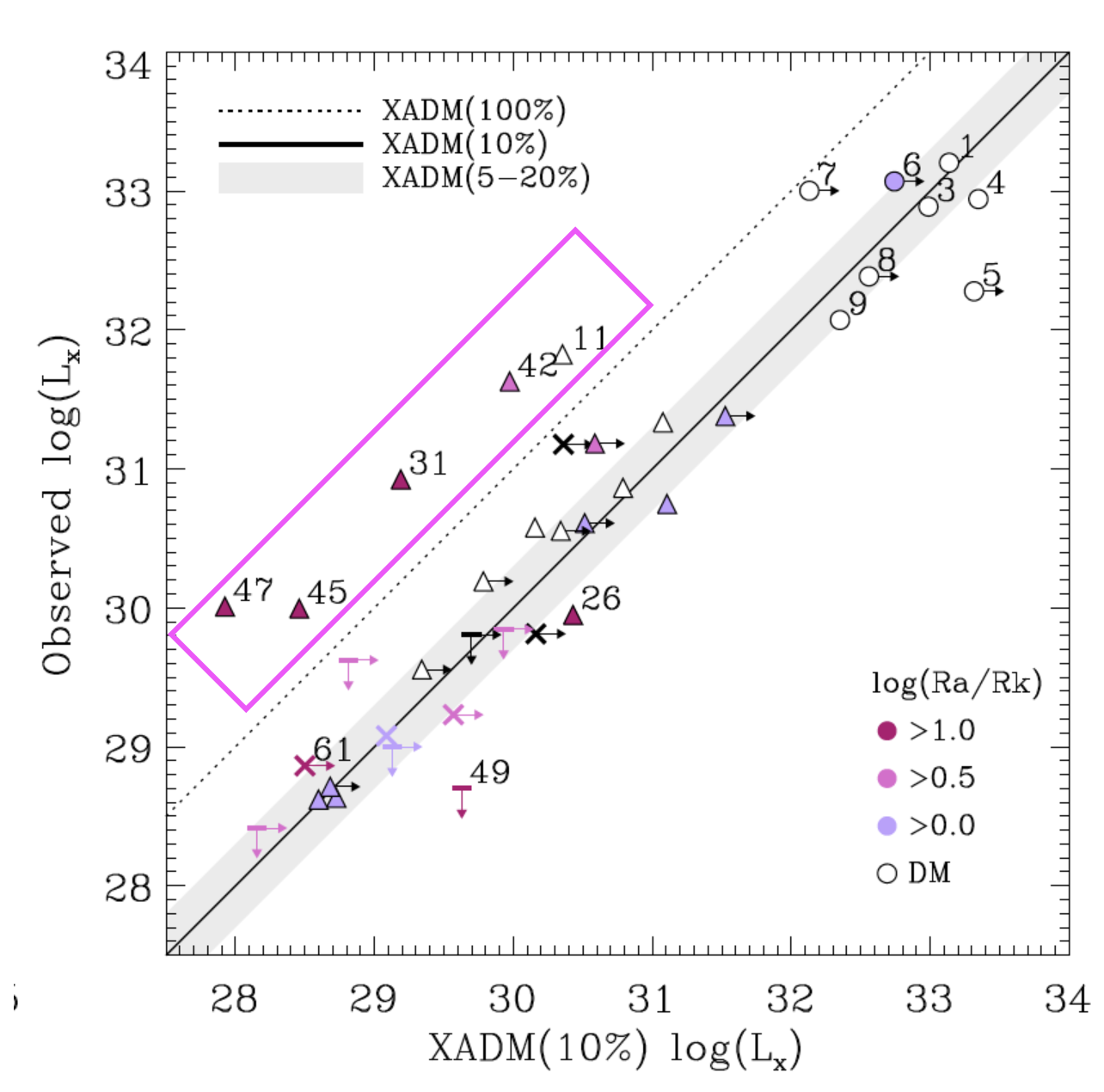}
\end{center}
\vspace{0.in}
\caption{ \small{
Upper left: Schematic showing ``shock retreat" that depends on the input mass flux into a closed magnetic loops.
Upper right: Comparison of resulting ``XADM" scaling relations (for X-rays from analytic dynamical magnetospheres) with time-averaged X-ray emission from 2D MHD simulations with 
$\eta_\ast =10$ (blue) and $\eta_\ast =100$ (black) (with bolder curves showing results when inverse-Compton cooling is also included). The dotted curves are for the XADM analytic scaling for the same models.
%(ud-Doula et al.\ 2014).
Lower left: Radius and time evolution of X-rays from MCWS in these 2D MHD simulations of DMs, showing the cycle of build up and decay, leading to the 20\% duty-cycle reduction compared with the XADM scalings.
Lower right:
Compilation \citep{Naze14} of  the observed X-ray luminosity of magnetic stars,  compared with the prediction  of the XADM model with a reduced 10\%  duty-cycle factor.
The agreement is best for slowly rotating DM stars with open circles (O-stars) and triangles (B-stars).
The stars within the magenta box, which show an order of magnitude higher X-ray emission, are almost all rapidly rotating B-stars with CMs (filled triangles).
}
}
\label{fig:xrays}
\end{figure}

\subsection{X-ray luminosity from magnetically confined wind shocks}
\label{sec:xrays}

For the DM cases without dynamically significant rotation,  i.e. small values of $W$,  \cite{Uddoula14} carried out a more systematic 2D MHD simulation parameter study with a full energy equation to compute the X-ray luminosity $L_x$ that results from { magnetically confined wind shocks}.
These MHD simulation results  were  used to calibrate an `X-ray Analytic Dynamical Magnetosphere' (XADM)  analysis for how the overall X-ray luminosity scales with stellar magnetic field strength and wind mass loss rate.

Figure \ref{fig:xrays} shows some key results, together with comparison with observed X-ray data.
The X-ray emission is largely controlled by a shock retreat effect (upper left panel) that depends on the wind feeding rate, and the associated radiative cooling length.
The upper-right panel plots the ratio $L_x/L_{bol}$ vs. wind feeding rate ${\dot M}$ for models with magnetic confinement $\etas= 10$ (blue) and $100$ (black), compared with prediction scalings (dashed curves) from the XADM analysis.
 For the most luminous stars, $L_x$ scales in proportion to the wind mass loss rate, but at lower $L_{bol}$, the lower $\Mdot$ means the radiative cooling becomes inefficient, leading to shock retreat such that weaker and softer X-rays are produced.
The lower-left panel showing the radius and time evolution of the X-ray emission exhibits episodes of little emission associated with intervals of matter infall, without shock heating.
This overall duty-cycle lowers  the averaged emission frequency relative to the idealized XADM analysis.

The lower right panel of figure \ref{fig:xrays}  shows that the XADM model with a 10\% efficiency matches quite well the trend in observed X-rays for a large sample of DM stars, over more that four decades in X-ray flux.
Stars in the magenta box, which are X-ray over-luminous relative to this XADM scaling, are in fact mostly rapid rotators with CMs.
This could possibly be  explained, e.g., by enhanced centrifugal acceleration that leads to stronger shocks, and thus harder, more intense X-ray emission; or, perhaps, by centrifugal (CBO) events and the associated heating by magnetic reconnection to X-ray emitting temperatures. For a more definitive answer, further future study must be undertaken.

\subsection{UV wind line variation observed by {\tt HST}}
\label{sec:UVHST}

The magnetic channeling of a hot-star wind outflow can also impart distinct signatures on UV wind lines that  can be observed with {\tt Hubble Space Telescope} (HST) \citep{David-Uraz19}.
In particular, multiple time exposures of slowly rotating magnetic O-stars show variations in the absorption troughs of UV P-Cygni lines like SiIV and CIV, with the most distinctive signatures seen in the O7f?cp star NGC 1624-2, the most strongly magnetic O-star  known \citep{David-Uraz21}.

\begin{figure}[t!]
	\begin{center}
	\includegraphics[width=115mm]{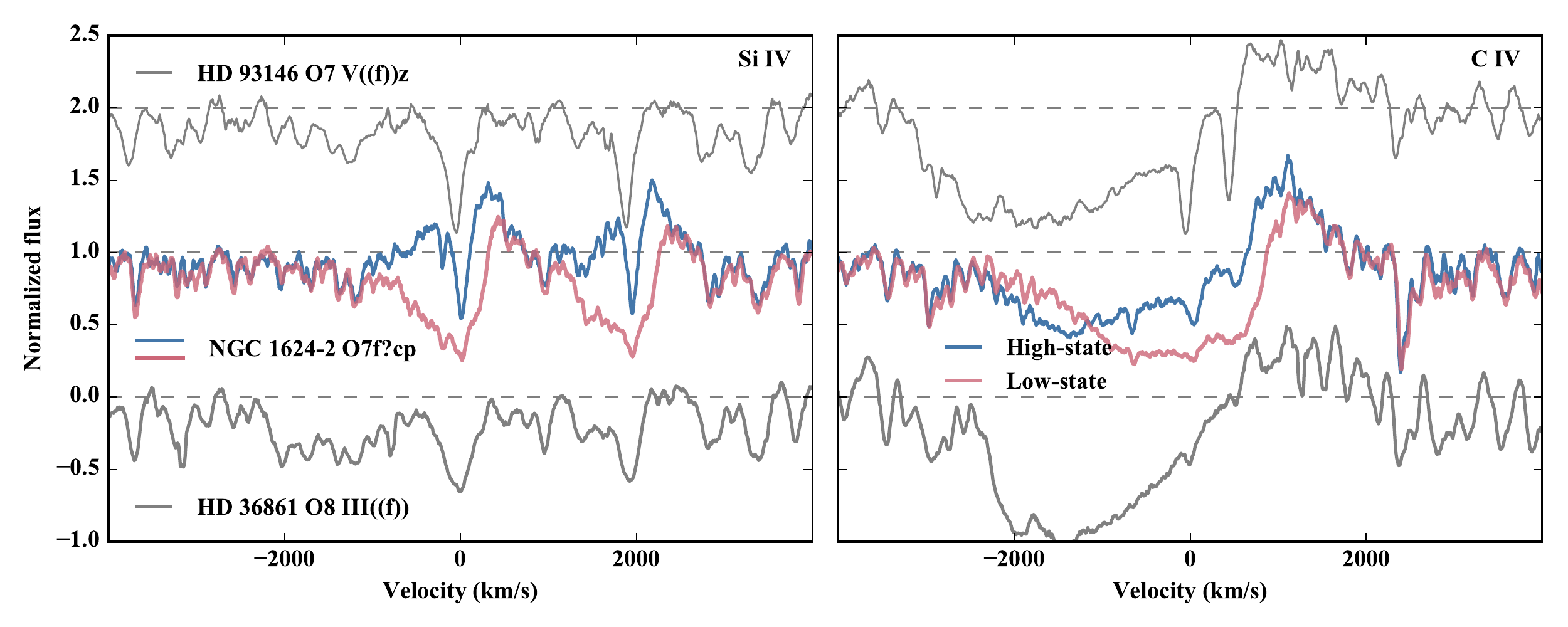}
	\includegraphics[width=58mm]{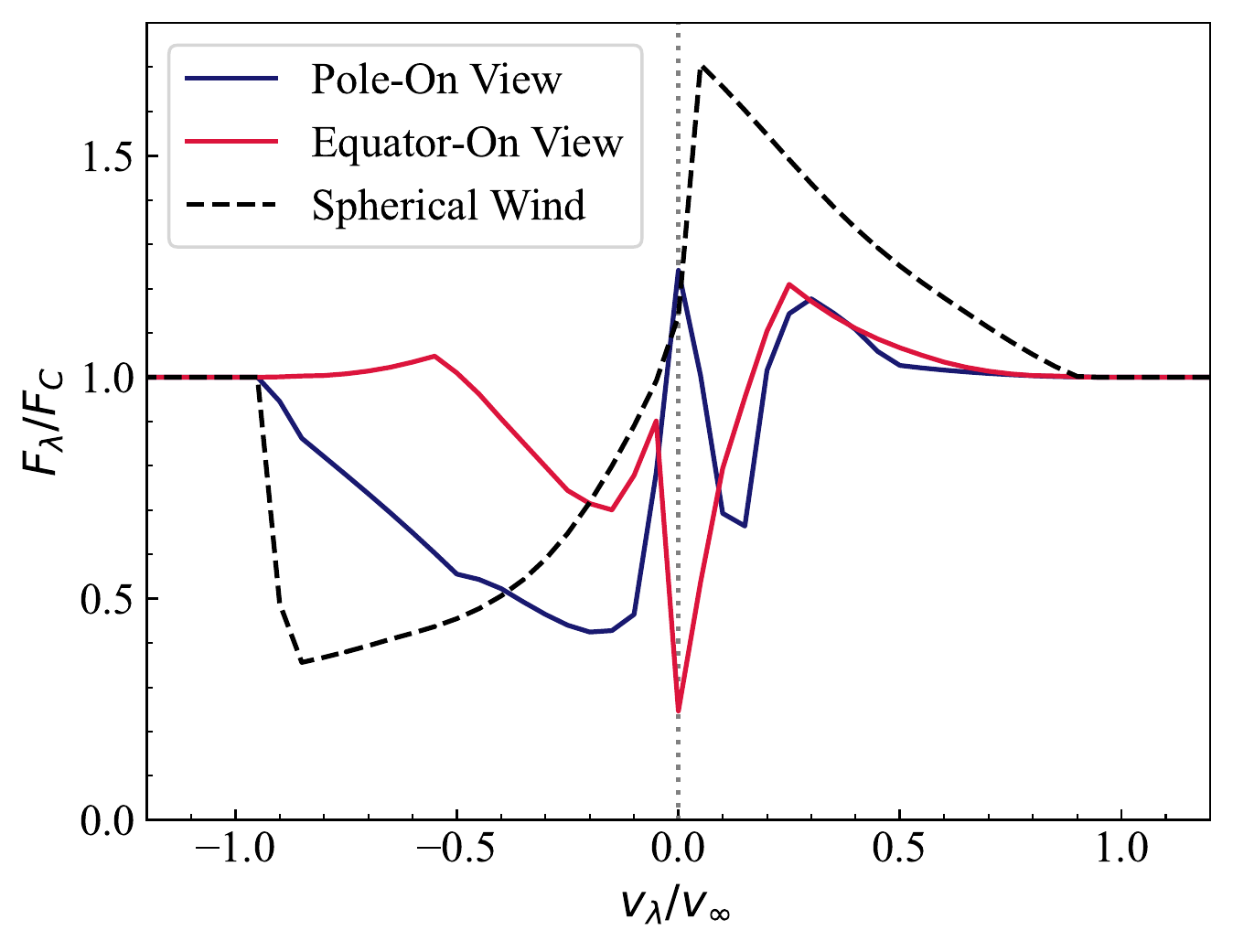}
	\includegraphics[width=58mm]{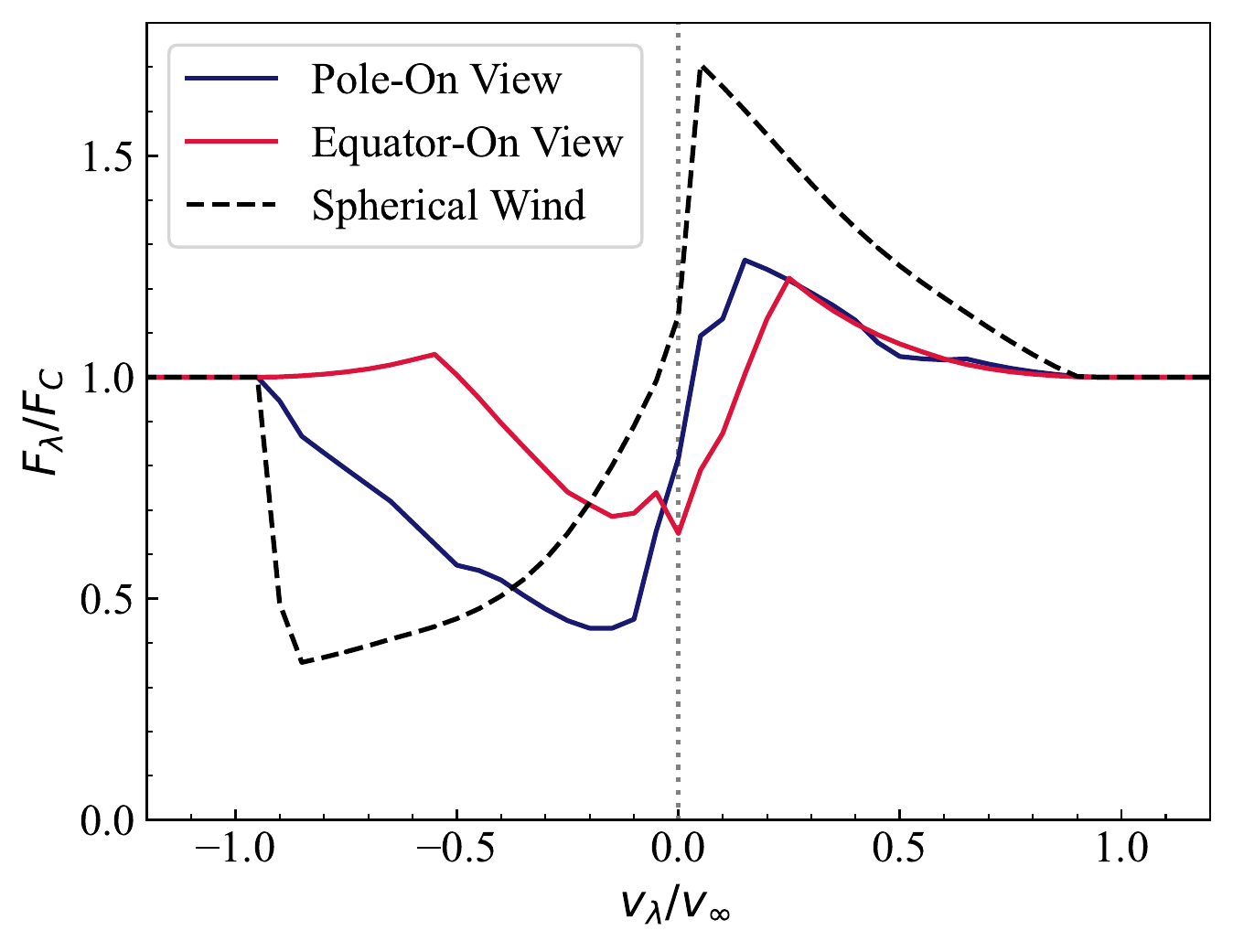}

	\caption{\small{
Top: {\tt HST} spectra of  Si IV (left panel) and C IV (right panel)  from two non-magnetic O-star (HD 93146, top spectrum, and HD 36861, bottom spectrum) compared with separate exposures from the strongly magnetic O-star NGC1624-2, showing both a `high' and `low' state.
Bottom: Synthetic line profiles showing how these Si IV lines (left) and  C IV lines (right) can be modeled with the analytic dynamical magnetosphere (ADM) as arising from the difference between polar and equatorial views of a tilted dipole field \citep{Erba17,Erba19}.
%(Erba et al.\ 2017, 2021).
 }
 }
 \label{fig:dash2}
\end{center}
\end{figure}

This and other magnetic O-stars are too slowly rotating to  have CMs; but they are nonetheless  inferred to have field tilts that give  rotational changes in observer perspective, leading in NGC 1624-2 to variation from a `high state' (with deeper, faster absorption) and a `low state' (with shallower, slower absorption).
Figure \ref{fig:dash2} shows how such profiles  can be modeled using the ADM formalism in terms of whether the observer views along the magnetic pole or equator of a tilted dipole (\citep{Erba17,Erba19}.
Similar spectral line-profile variations are seen in multiple {\tt HST} exposures from other magnetic O-stars \citep{David-Uraz19}.

\begin{figure}[t!]
{
	\includegraphics[width=115mm]{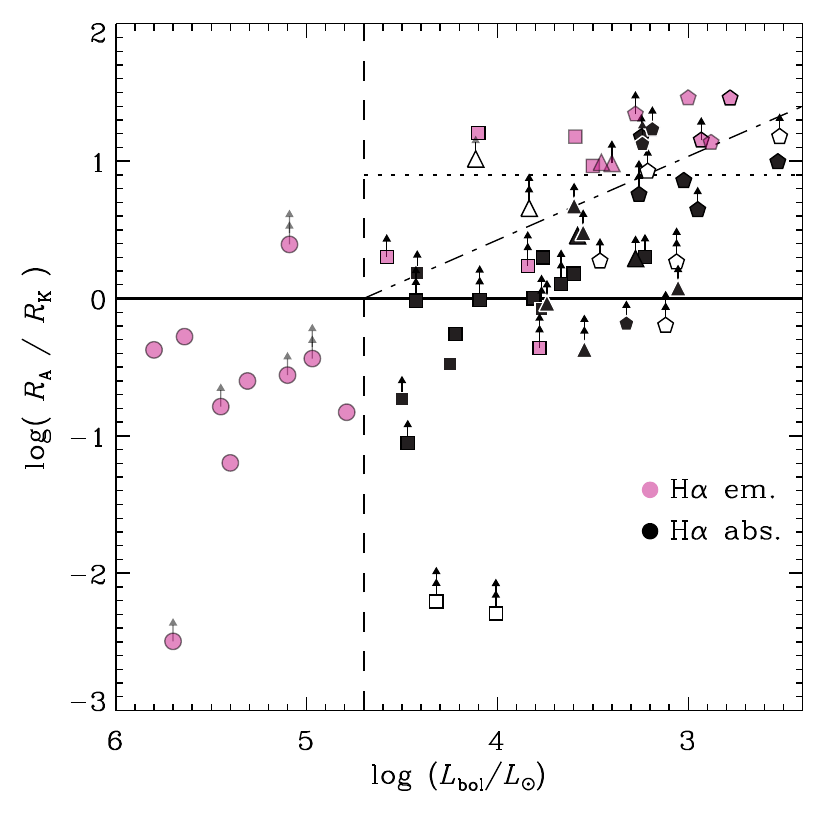}
}
\caption{\small{
Location of magnetic massive stars in a $\log$-$\log$ plot of $\ra/\rk$ vs.\ stellar luminosity. 
The symbol shadings mark the presence (pink or shaded) or absence (black) of magnetospheric H$\alpha$ emission.
The main text discusses how centrifgual breakout (CBO) explains this onset of H$\alpha$ emission.
 }
 }
\label{fig:ha}
\end{figure}

\begin{figure}[t!]
	\begin{center}
	\includegraphics[scale=.43]{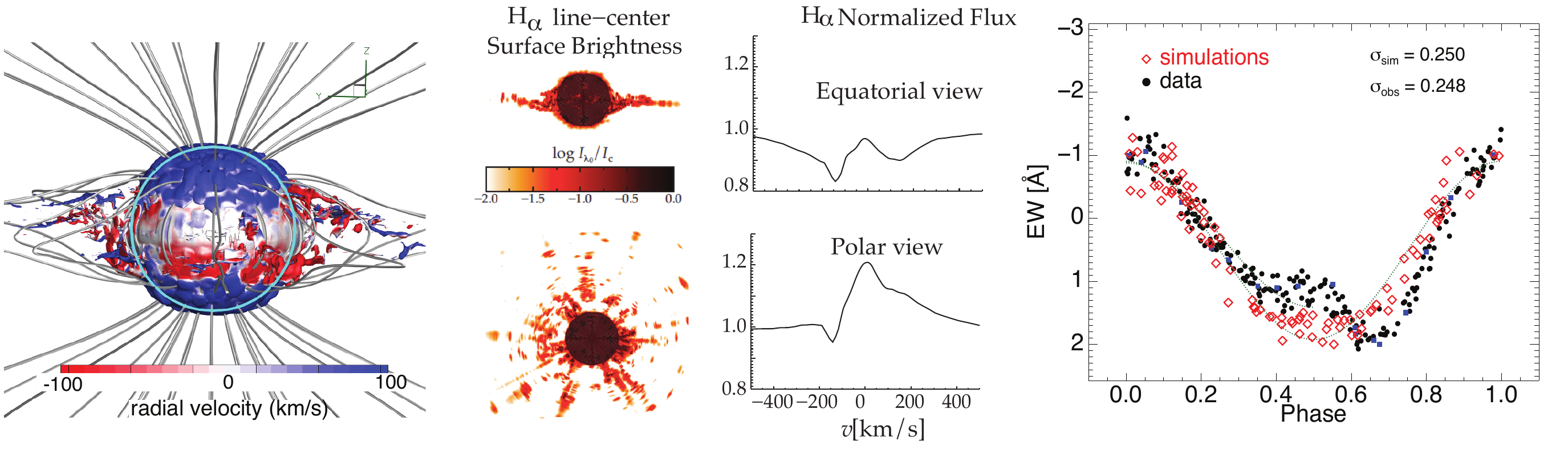}
	\caption{\small{
3D MHD model of the dynamical magnetosphere for the young, slowly rotating (15.4-day period) O7V star \thooc  \citep{Uddoula13}.
%(ud-Doula et al.\ 2013).
The left panel shows a snapshot of wind structure drawn as isodensity surface,
colored to show radial component of velocity.
The middle panels show the predicted equatorial and polar views of H$\alpha$ line-center surface brightness, along with corresponding line-flux profiles.
The right panel compares the observed rotational modulation of the H$\alpha$ equivalent width (black) with 3D model predictions (red) assuming a pure-dipole surface field tilted by $\beta = 45^\circ$ to the rotation axis, as viewed from the inferred observer inclination of $i = 45^\circ$.
 }
 }
 \label{fig:3DT1OC}
\end{center}
\end{figure}

\subsection{H$\alpha$ line emission from Dynamical Magnetospheres}
\label{sec:HaDM}

The figure \ref{fig:ha} plots the observed magnetic stars in a plane comparing the  ratio $\Ralf/\Rkep$ vs.\ stellar luminosity, with now the symbol coded to mark the presence (light shading) or absence (black) of magnetospheric H$\alpha$ emission.
The horizontal solid line marks the transition between the CM domain above and the DM domain below, while the vertical dashed line marks the observational divide between O- and B-type  main sequence stars.
Note that {\em all} O-stars show H$\alpha$ emission, even though they are  located among the slow rotators with a DM.
By contrast,  most B-type stars only show H$\alpha$ emission if they are well above the $\Ralf/\Rkep = 1$ horizontal line, implying a relatively fast rotation and strong confinement that leads to a CM.

The basic explanation for this dichotomy is straightforward. The stronger winds driven by the higher luminosity O-stars can accumulate even within a relatively short dynamical timescale to a sufficient density to give the strong H$\alpha$ emission in a DM, while the weaker winds of lower luminosity B-stars require the longer confinement and buildup of a CM to reach densities required for such emission.
This general picture is confirmed by the detailed  dynamical models of DM and CM H$\alpha$  emission that motivated this empirical classification \cite{udD2009, Sundqvist12}.

For  the slowly rotating  O-stars HD\,191612 and \thooc\, 
the relative unimportance of centrifugal effects allows
both 2D and 3D MHD simulations \citep{Sundqvist12,Uddoula13}
 %(Sundqvist et al.\ 2012; ud-Doula et al.\ 2013)
of the wind-fed DM, reproducing quite well the rotational variation of H$\alpha$ emission.
For the 3D simulations of \thooc, figure \ref{fig:3DT1OC} shows how wind material trapped in closed loops over the magnetic equator (left panel) leads to circumstellar emission that is strongest during rotational phases corresponding to pole-on views (middle panel). For a pure dipole with the inferred magnetic tilt  $\beta=45^\circ$, an observer with  the inferred inclination $i=45^\circ$ has  perspectives that vary from magnetic pole to equator, leading in the 3D model to the rotational phase variations in H$\alpha$ equivalent width shown in the right panel (red squares). This matches quite well both the modulation and random fluctuation of the observed equivalent width (black dots).

\begin{figure}[t!]
\vspace{-0.0in}
\begin{center}
\includegraphics[scale=0.42]{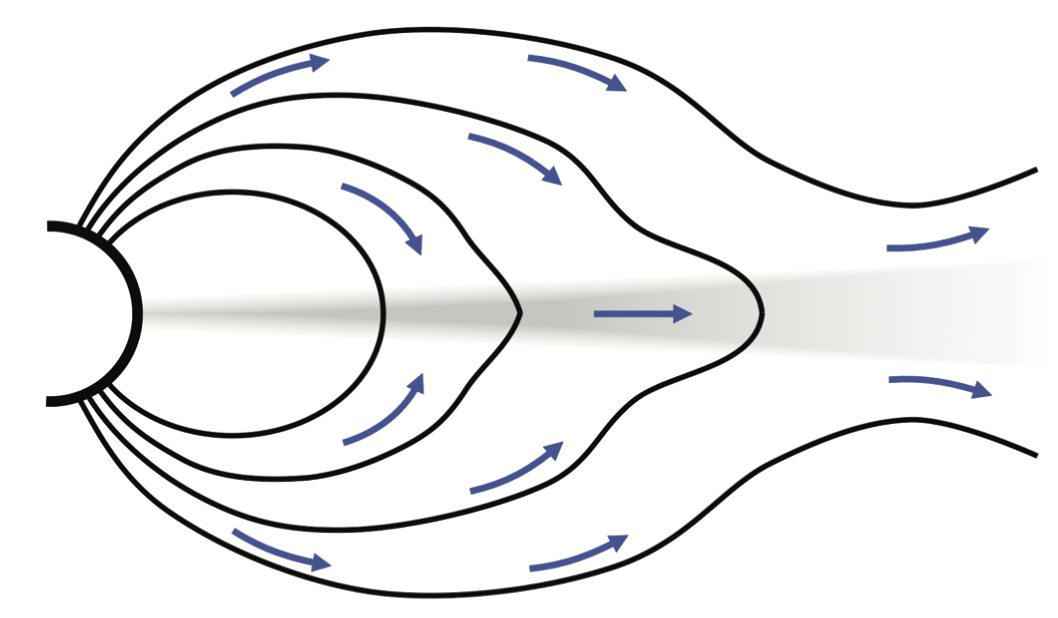}
\includegraphics[scale=0.337]{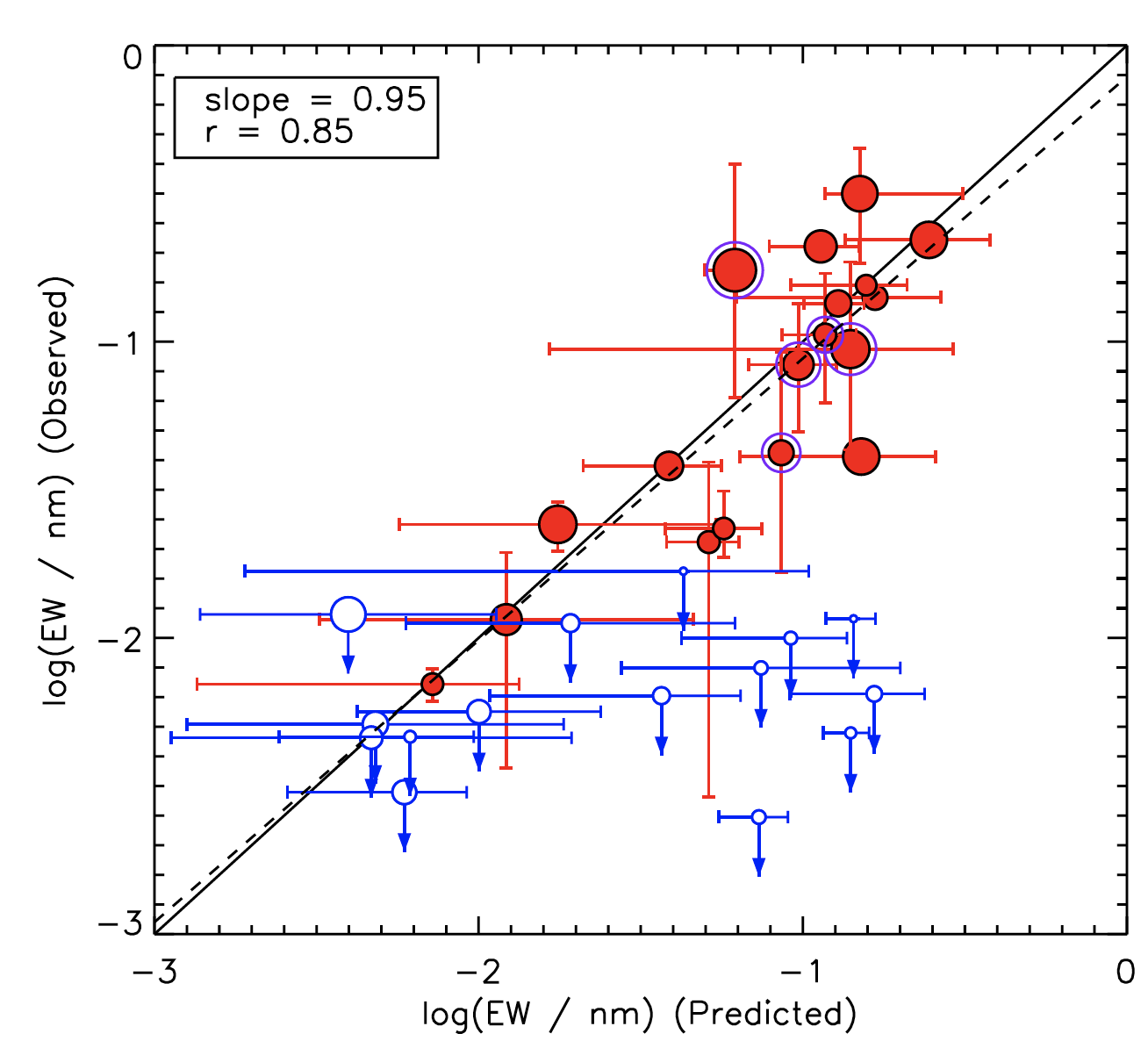}
\includegraphics[scale=0.353]{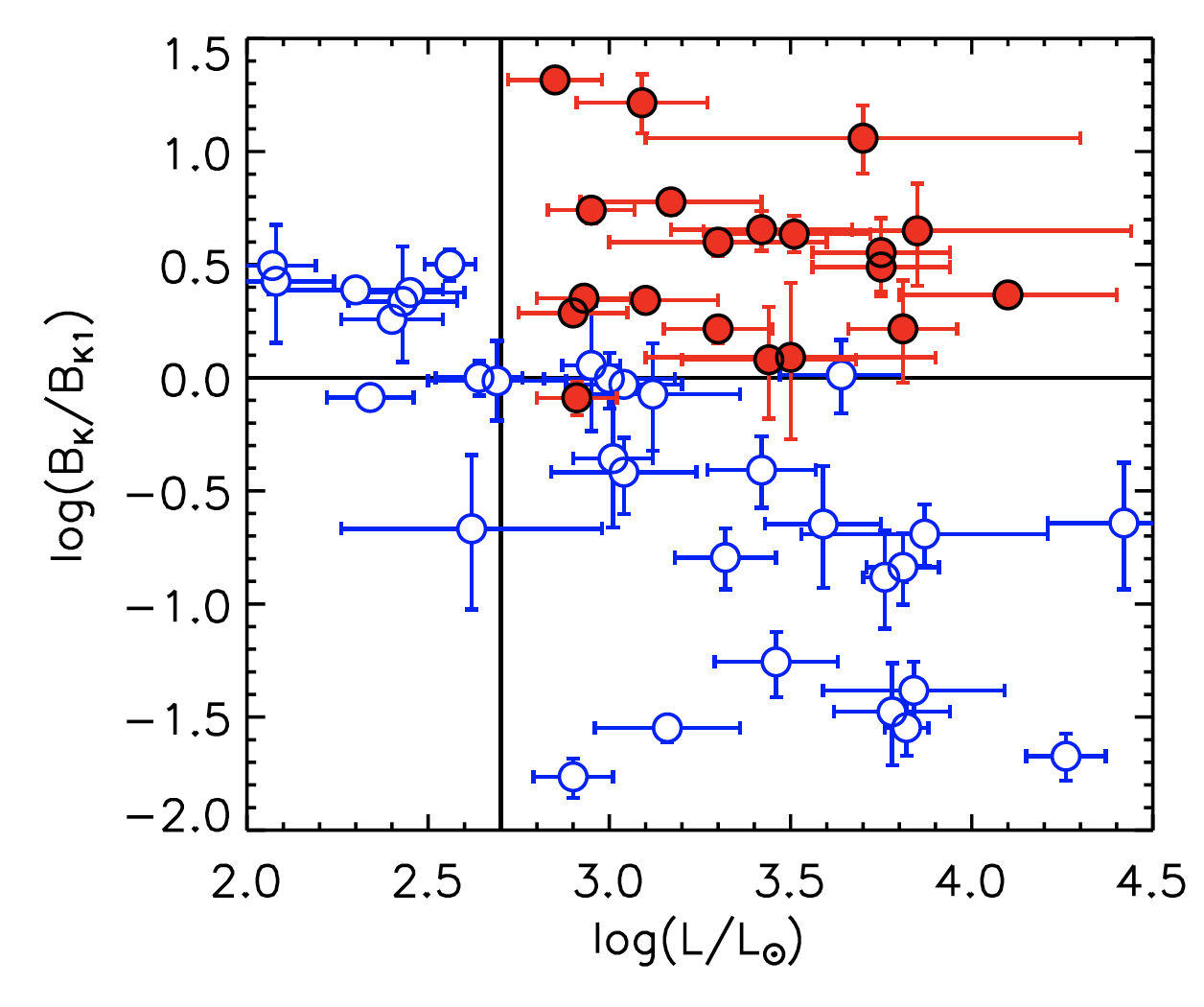}
\end{center}
\vspace{-0.in}
\caption{\small{
Top: Schematic to show the process of centrifugal breakout (CBO).
Center: For the magnetic B-star sample from \cite{Shultz21}, measured equivalent width (EW) of H$\alpha$ emission vs.\ predicted values from the CBO model.
Bottom: The same sample as above plotted as $\log B_K/B_{K1}$ vs. $\log L/L_\odot$, with stars showing H$\alpha$ emission in filled red circles, and those without in open blue circles.
Here $B_K$ is the observationally inferred field strength at the Kepler co-rotation radius $\rk$, while  $B_{K1}$ is the strength the CBO predicts is needed to make the CM have unit optical thickness in H$\alpha$ at $\rk$.
Note the remarkable agreement of the CBO predictions for both the onset and level of H$\alpha$ emission.
(Low luminosity stars to the left of the vertical line may have too weak a wind to fill the CM over other competing leakage mechanisms.)}
}
\label{fig:cbo}
\end{figure}

\subsection{Centrifugal Breakout and H$\alpha$ emission from Centrifugal Magnetospheres}
\label{sec:CBO}

In contrast to the dynamical cycle of upflow and infall in DMs, for  CMs it is less clear how the mass buildup from the trapping of material between $\Rkep$ and $\Ralf$ is to be balanced.  
One possibility is that  outward drift and diffusion from turbulent cross-field transport allows a gradual, quasi-steady leakage that balances the wind-feeding after a certain level of buildup, which depends on the mass filling rate and the magnitude of the turbulent transport coefficients \citep{Owocki18}.
%(Owocki and Cranmer 2018).

But recent studies of H$\alpha$ emission from CMs  \citep{Shultz20,Owocki20} seem to strongly favor a CBO paradigm, in which magnetic confinement of mass in the co-rotating CM builds up until the centrifugal force overwhelms the magnetic tension, leading then to sporadic episodes of CBO, e.g., as seen in the $\Delta m/\Delta r$ plots of figure \ref{fig:dmdr}.

Figure \ref{fig:cbo} shows that this CBO model explains well  both the onset and equivalent width of H$\alpha$ emission.
In the bottom panel, $B_{K1}$ represents a prediction of the CBO analysis for the magnetic field strength at the Kepler radius needed to make the CM have unit optical depth in H$\alpha$ there;
note that only B-stars with $B_{K} > B_{K1}$ exhibit emission, showing its onset occurs just at the field strength needed to confine material to the density to make the CM optically thick in H$\alpha$.
The center panel shows further a very good agreement between the observed vs. CBO-predicted equivalent width of H$\alpha$.

Moreover,  \cite{Shultz21}  %(Shultz et al.\ 2021) 
indicate that detections of non-thermal radio emission from the magnetic B-stars lie in a very similar region of the rotation-confinement diagram as H$\alpha$, implying that it too depends on a combination of strong confinement and rapid rotation. Models currently being developed indicate this too may be explained by CBO, through acceleration of electrons in CBO-driven magnetic reconnection, followed by their gyro-emission in the magnetic field.

\subsection{CBO challenges to rigid-field models}
\label{sec:CBO-RRM}

\begin{figure}[t!]
\vspace{-0.0in}
\begin{center}
\includegraphics[scale=0.22]{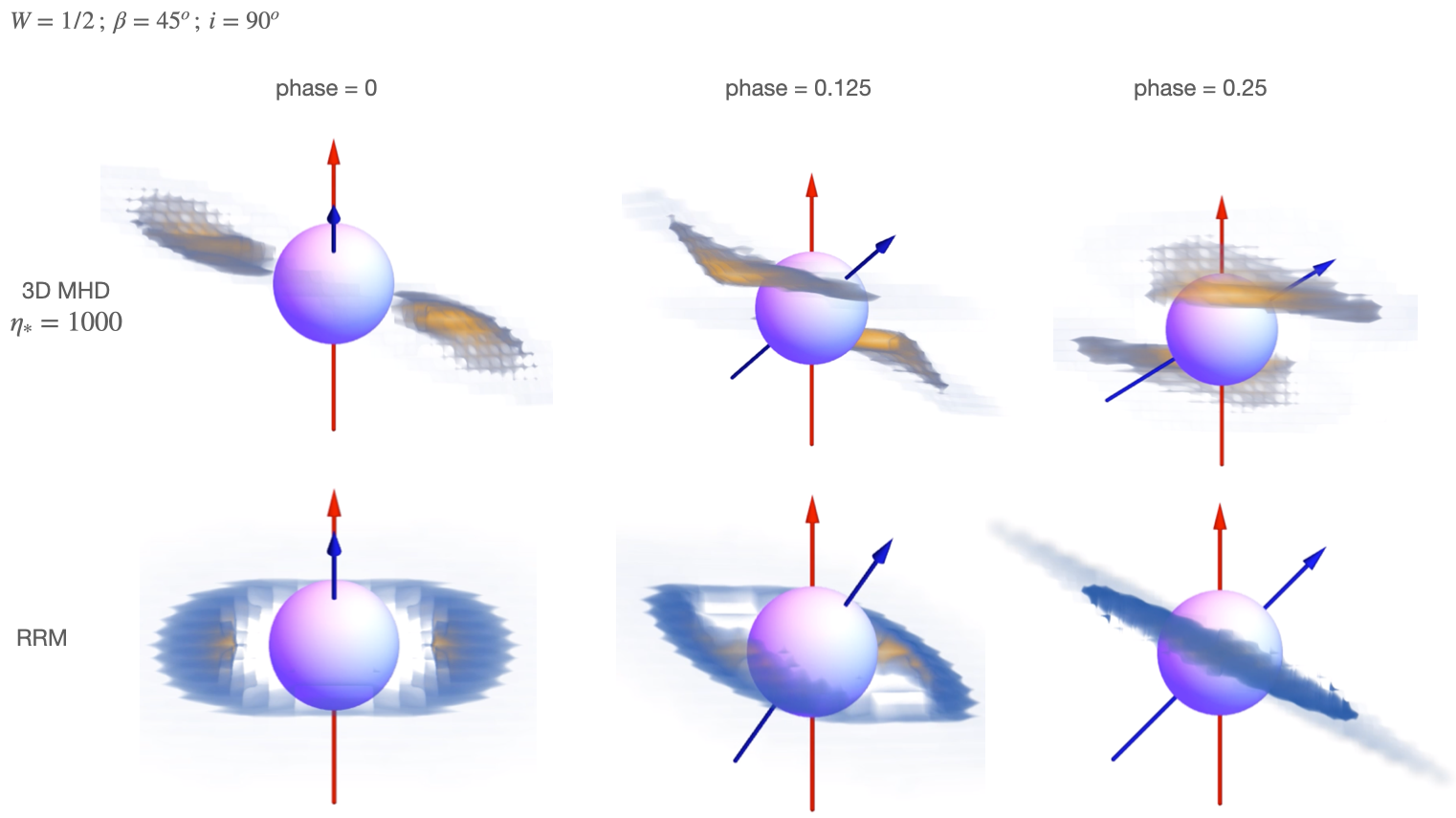}
\end{center}
\vspace{-0.in}
\caption{ \small{
For for a tilted dipole case with W=1/2 and $\beta=45^\circ$,
a direct comparison of the RRM model and a preliminary 3D MHD simulation, done with a 3D implementation of the \textsc{Pluto} MHD code, here for a case with large magnetic confinement,  $\eta_\ast = 1000$. The plot shows column density  viewed from inclination $i=90^\circ$ at 3 representative rotational phases.
Note the significant differences in both azimuthal and latitudinal distribution of mass, which will lead to essential differences in observational signatures.
}
}
\label{fig:mhd-vs-rrm}
\end{figure}

These recent results favoring CBO present a challenge to previous attempts to use an idealization of a purely {\em rigid-field} to model CMs.
The Rigidly Rotating Magnetosphere \citep[RRM;][] {Tow2005} model provides a semi-analytical prescription for the 3D magnetospheric plasma distribution, based on the form and minima of the total gravitational-plus-centrifugal potential along each separate field line.  
The Rigid-Field Hydrodynamics \citep[RFHD;][] {Tow2008} approach carries out separate hydrodynamical simulations of wind flow on a large number ($>1000$) of rigid field lines, to derive both the shocked feeding and final settlement of material at such potential minima.
Both approaches have provided an instructive basis for 3D models of CM's with tilted dipoles or even complex fields, along with associated observational diagnostics like H$\alpha$ and X-ray emission \citep{Tow2008,Oksala2012}.
%(Townsend et al.\ 2006; Oksala et al.\ 2012).

However, because they provide no mechanisms for {\em emptying} the mass build-up from wind material trapped in the CM above the Kepler radius,
the resulting mass distribution is simply set by the local wind feeding rate, with a value that increases secularly with an arbitrarily assumed `filling time'.

Moreover, while the rigid-field assumption seems well justified in guiding the wind outflow -- for which the associated confinement parameters can be as high as $\eta_\ast = 10^6$ --,
the notion that CBO controls the ultimate mass escape that balances the wind feeding implies that the field in such CMs must be under active dynamic stress, and so no longer acting as a purely rigid guide.
The net effects of such CBO distortions on the overall 3D plasma distribution are thus not yet understood and further investigations are necessary. 

Figure \ref{fig:mhd-vs-rrm} shows some preliminary results comparing 3D MHD simulations (using a 3D version of \textsc{Pluto}) with the RRM model.
Understanding the reasons for the clear differences will be a central  focus of a future study.

\section{Future Outlook}
Magnetic massive stars are sources of  hard X-ray  emission. To understand fully how MCWS paradigm works in real stars, we need improved 3D MHD modelling that includes full energy equation with an appropriate radiative cooling. Most current 3D MHD models use isothermal equation of state whereas the RRM model lacks the necessary wind dynamics. In addition, we need to have a better understanding what role rotation plays in magnetic massive stars. The CBO paradigm provides us with a clear clue, but it does not provide us with a detailed mechanism how such continuous small scale breakout events take place.

\section{Cross-References}
For the observational aspects of MCWS, please see the chapter by Gregor Rauw.

%\subsection{References}

\bibliography{ud-Doula}

\begin{thebibliography}{10}

\bibitem{Chlebowski89}
T.~{Chlebowski}, F.~R. {Harnden}, Jr., and S.~{Sciortino}, ``{The Einstein
  X-ray Observatory Catalog of O-type stars},'' {\em \apj}, vol.~341,
  pp.~427--455, June 1989.

\bibitem{Berghoefer97}
T.~W. {Berghoefer}, J.~H.~M.~M. {Schmitt}, R.~{Danner}, and J.~P. {Cassinelli},
  ``{X-ray properties of bright OB-type stars detected in the ROSAT all-sky
  survey.},'' {\em \aap}, vol.~322, pp.~167--174, June 1997.

\bibitem{Naze11}
Y.~{Naz{\'e}}, P.~S. {Broos}, L.~{Oskinova}, L.~K. {Townsley}, D.~{Cohen},
  M.~F. {Corcoran}, N.~R. {Evans}, M.~{Gagn{\'e}}, A.~F.~J. {Moffat}, J.~M.
  {Pittard}, G.~{Rauw}, A.~{ud-Doula}, and N.~R. {Walborn}, ``{Global X-ray
  Properties of the O and B Stars in Carina},'' {\em \apjs}, vol.~194, p.~7,
  May 2011.

\bibitem{Petit13}
V.~{Petit}, S.~P. {Owocki}, G.~A. {Wade}, D.~H. {Cohen}, J.~O. {Sundqvist},
  M.~{Gagn{\'e}}, J.~{Ma{\'{\i}}z Apell{\'a}niz}, M.~E. {Oksala}, D.~A.
  {Bohlender}, T.~{Rivinius}, H.~F. {Henrichs}, E.~{Alecian}, R.~H.~D.
  {Townsend}, A.~{ud-Doula}, and {MiMeS Collaboration}, ``{A magnetic
  confinement versus rotation classification of massive-star magnetospheres},''
  {\em \mnras}, vol.~429, pp.~398--422, Feb. 2013.

\bibitem{BabMon1997a}
J.~{Babel} and T.~{Montmerle}, ``{X-ray emission from Ap-Bp stars: a
  magnetically confined wind-shock model for IQ Aur.},'' {\em \aap}, vol.~323,
  pp.~121--138, July 1997.

\bibitem{BabMon1997b}
J.~{Babel} and T.~{Montmerle}, ``{On the Periodic X-Ray Emission from the O7 V
  Star theta 1 Orionis C},'' {\em \apjl}, vol.~485, p.~L29, Aug. 1997.

\bibitem{Gag2005}
M.~{Gagn{\'e}}, M.~E. {Oksala}, D.~H. {Cohen}, S.~K. {Tonnesen}, A.~{ud-Doula},
  S.~P. {Owocki}, R.~H.~D. {Townsend}, and J.~J. {MacFarlane}, ``{Chandra HETGS
  Multiphase Spectroscopy of the Young Magnetic O Star ${\theta}^{1}$ Orionis
  C},'' {\em \apj}, vol.~628, pp.~986--1005, Aug. 2005.

\bibitem{ShoBro1990}
S.~N. {Shore} and D.~N. {Brown}, ``{Magnetically controlled circumstellar
  matter in the helium-strong stars},'' {\em \apj}, vol.~365, pp.~665--676,
  Dec. 1990.

\bibitem{udDOwo2002}
A.~{ud-Doula} and S.~P. {Owocki}, ``{Dynamical Simulations of Magnetically
  Channeled Line-driven Stellar Winds. I. Isothermal, Nonrotating, Radially
  Driven Flow},'' {\em \apj}, vol.~576, pp.~413--428, Sept. 2002.

\bibitem{Uddoula08}
A.~{ud-Doula}, S.~P. {Owocki}, and R.~H.~D. {Townsend}, ``{Dynamical
  simulations of magnetically channelled line-driven stellar winds - II. The
  effects of field-aligned rotation},'' {\em \mnras}, vol.~385, pp.~97--108,
  Mar. 2008.

\bibitem{Uddoula14}
A.~{ud-Doula}, S.~{Owocki}, R.~{Townsend}, V.~{Petit}, and D.~{Cohen},
  ``{X-rays from magnetically confined wind shocks: effect of cooling-regulated
  shock retreat},'' {\em \mnras}, vol.~441, pp.~3600--3614, July 2014.

\bibitem{udD2003}
A.~{ud-Doula}, {\em {The effects of magnetic fields and field-aligned rotation
  on line-driven hot-star winds}}.
\newblock PhD thesis, University of Delaware, Feb. 2003.

\bibitem{Petit17}
V.~{Petit}, Z.~{Keszthelyi}, R.~{MacInnis}, D.~H. {Cohen}, R.~H.~D. {Townsend},
  G.~A. {Wade}, S.~L. {Thomas}, S.~P. {Owocki}, J.~{Puls}, and A.~{ud-Doula},
  ``{Magnetic massive stars as progenitors of `heavy' stellar-mass black
  holes},'' {\em \mnras}, vol.~466, pp.~1052--1060, Apr. 2017.

\bibitem{Owocki16}
S.~P. {Owocki}, A.~{ud-Doula}, J.~O. {Sundqvist}, V.~{Petit}, D.~H. {Cohen},
  and R.~H.~D. {Townsend}, ``{An `analytic dynamical magnetosphere' formalism
  for X-ray and optical emission from slowly rotating magnetic massive
  stars},'' {\em \mnras}, vol.~462, pp.~3830--3844, Nov. 2016.

\bibitem{Erba19}
C.~{Erba}, A.~{David-Uraz}, and V.~{Petit}, ``{Quantitative Modeling of the
  Ultraviolet Spectra of Magnetic Massive Stars}.'' HST Proposal, June 2019.

\bibitem{Sundqvist12}
J.~O. {Sundqvist} and S.~P. {Owocki}, ``{Clumping in the inner winds of hot,
  massive stars from hydrodynamical line-driven instability simulations},''
  {\em \mnras}, p.~144, Nov. 2012.

\bibitem{Uddoula09}
A.~{ud-Doula}, S.~P. {Owocki}, and R.~H.~D. {Townsend}, ``{Dynamical
  simulations of magnetically channelled line-driven stellar winds - III.
  Angular momentum loss and rotational spin-down},'' {\em \mnras}, vol.~392,
  pp.~1022--1033, Jan. 2009.

\bibitem{Weber67}
E.~J. {Weber} and J.~{Davis}, Leverett, ``{The Angular Momentum of the Solar
  Wind},'' {\em \apj}, vol.~148, pp.~217--227, Apr. 1967.

\bibitem{Keszthelyi20}
Z.~{Keszthelyi}, G.~{Meynet}, M.~E. {Shultz}, A.~{David-Uraz}, A.~{ud-Doula},
  R.~H.~D. {Townsend}, G.~A. {Wade}, C.~{Georgy}, V.~{Petit}, and S.~P.
  {Owocki}, ``{The effects of surface fossil magnetic fields on massive star
  evolution - II. Implementation of magnetic braking in MESA and implications
  for the evolution of surface rotation in OB stars},'' {\em \mnras}, vol.~493,
  pp.~518--535, Jan. 2020.

\bibitem{Tow2010}
R.~H.~D. {Townsend}, M.~E. {Oksala}, D.~H. {Cohen}, S.~P. {Owocki}, and
  A.~{ud-Doula}, ``{Discovery of Rotational Braking in the Magnetic
  Helium-strong Star Sigma Orionis E},'' {\em \apjl}, vol.~714, pp.~L318--L322,
  May 2010.

\bibitem{Naze14}
Y.~{Naz{\'e}}, V.~{Petit}, M.~{Rinbrand}, D.~{Cohen}, S.~{Owocki},
  A.~{ud-Doula}, and G.~A. {Wade}, ``{X-Ray Emission from Magnetic Massive
  Stars},'' {\em \apjs}, vol.~215, p.~10, Nov. 2014.

\bibitem{David-Uraz19}
A.~{David-Uraz}, C.~{Erba}, V.~{Petit}, A.~W. {Fullerton}, F.~{Martins}, N.~R.
  {Walborn}, R.~{MacInnis}, R.~H. {Barb{\'a}}, D.~H. {Cohen}, J.~{Ma{\'\i}z
  Apell{\'a}niz}, Y.~{Naz{\'e}}, S.~P. {Owocki}, J.~O. {Sundqvist},
  A.~{ud-Doula}, and G.~A. {Wade}, ``{Extreme resonance line profile variations
  in the ultraviolet spectra of NGC 1624-2: probing the giant magnetosphere of
  the most strongly magnetized known O-type star},'' {\em \mnras}, vol.~483,
  pp.~2814--2824, Feb. 2019.

\bibitem{David-Uraz21}
A.~{David-Uraz}, V.~{Petit}, M.~E. {Shultz}, A.~W. {Fullerton}, C.~{Erba},
  Z.~{Keszthelyi}, S.~{Seadrow}, and G.~A. {Wade}, ``{New observations of NGC
  1624-2 reveal a complex magnetospheric structure and underlying surface
  magnetic geometry},'' {\em \mnras}, vol.~501, pp.~2677--2687, Feb. 2021.

\bibitem{Erba17}
C.~{Erba}, A.~{David-Uraz}, V.~{Petit}, and S.~P. {Owocki}, ``{New Insights
  into the Puzzling P-Cygni Profiles of Magnetic Massive Stars},'' in {\em The
  Lives and Death-Throes of Massive Stars} (J.~J. {Eldridge}, J.~C. {Bray},
  L.~A.~S. {McClelland}, and L.~{Xiao}, eds.), vol.~329, pp.~246--249, Nov.
  2017.

\bibitem{Uddoula13}
A.~{ud-Doula}, J.~O. {Sundqvist}, S.~P. {Owocki}, V.~{Petit}, and R.~H.~D.
  {Townsend}

\bibitem{udD2009}
A.~{ud-Doula}, S.~P. {Owocki}, and R.~H.~D. {Townsend}, ``{Dynamical
  simulations of magnetically channelled line-driven stellar winds - III.
  Angular momentum loss and rotational spin-down},'' {\em \mnras}, vol.~392,
  pp.~1022--1033, Jan. 2009.

\bibitem{Owocki18}
S.~P. {Owocki} and S.~R. {Cranmer}, ``{Diffusion-plus-drift models for the mass
  leakage from centrifugal magnetospheres of magnetic hot-stars},'' {\em
  \mnras}, vol.~474, pp.~3090--3100, Mar. 2018.

\bibitem{Shultz20}
M.~E. {Shultz}, S.~{Owocki}, T.~{Rivinius}, G.~A. {Wade}, C.~{Neiner},
  E.~{Alecian}, O.~{Kochukhov}, D.~{Bohlender}, A.~{ud-Doula}, J.~D.
  {Landstreet}, J.~{Sikora}, A.~{David-Uraz}, V.~{Petit},
  P.~{Cerraho{\u{g}}lu}, R.~{Fine}, G.~{Henson}, {MiMeS Collaboration}, and
  {BinaMIcS Collaboration}, ``{The magnetic early B-type stars - IV. Breakout
  or leakage? H {\ensuremath{\alpha}} emission as a diagnostic of plasma
  transport in centrifugal magnetospheres},'' {\em \mnras}, vol.~499,
  pp.~5379--5395, Dec. 2020.

\bibitem{Owocki20}
S.~P. {Owocki}, M.~E. {Shultz}, A.~{ud-Doula}, J.~O. {Sundqvist}, R.~H.~D.
  {Townsend}, and S.~R. {Cranmer}, ``{How the breakout-limited mass in B-star
  centrifugal magnetospheres controls their circumstellar H
  {\ensuremath{\alpha}} emission},'' {\em \mnras}, vol.~499, pp.~5366--5378,
  Dec. 2020.

\bibitem{Shultz21}
M.~E. {Shultz}, S.~{Owocki}, T.~{Rivinius}, G.~A. {Wade}, C.~{Neiner},
  E.~{Alecian}, O.~{Kochukhov}, D.~{Bohlender}, A.~{ud-Doula}, J.~D.
  {Landstreet}, J.~{Sikora}, A.~{David-Uraz}, V.~{Petit},
  P.~{Cerraho{\u{g}}lu}, R.~{Fine}, G.~{Henson}, G.~{Henson}, {MiMeS
  Collaboratio}, and {BinaMIcS Collaboration}, ``{The magnetic early B-type
  stars - IV. Breakout or leakage? H {\ensuremath{\alpha}} emission as a
  diagnostic of plasma transport in centrifugal magnetospheres},'' {\em
  \mnras}, vol.~499, pp.~5379--5395, Dec. 2020.

\bibitem{Tow2005}
R.~H.~D. {Townsend}, S.~P. {Owocki}, and D.~{Groote}, ``{The Rigidly Rotating
  Magnetosphere of {$\sigma$} Orionis E},'' {\em \apjl}, vol.~630,
  pp.~L81--L84, Sept. 2005.

\bibitem{Tow2008}
R.~H.~D. {Townsend}, ``{Exploring the photometric signatures of magnetospheres
  around helium-strong stars},'' {\em \mnras}, vol.~389, pp.~559--566, Sept.
  2008.

\bibitem{Oksala2012}
M.~E. {Oksala}, G.~A. {Wade}, R.~H.~D. {Townsend}, S.~P. {Owocki},
  O.~{Kochukhov}, C.~{Neiner}, E.~{Alecian}, and J.~{Grunhut}, ``{Revisiting
  the Rigidly Rotating Magnetosphere model for {\ensuremath{\sigma}} Ori E - I.
  Observations and data analysis},'' {\em \mnras}, vol.~419, pp.~959--970, Jan.
  2012.

\end{thebibliography}

\end{document}